\pdfoutput=1
\documentclass[a4paper,12pt]{article}
\usepackage[a4paper, top=1.6cm, bottom=1.6cm, left=1.6cm, right=1.6cm]%
{geometry}
\usepackage[utf8]{inputenc}
\usepackage{setspace}
\usepackage{epsfig,epsf,psfrag, url, setspace, graphicx,mathtools,subfigure,comment}
\usepackage{a4wide,amsmath, amsthm, amssymb, multirow}
\usepackage{xcolor,colortbl}
\usepackage{verbatim}
\usepackage{arydshln} 
\usepackage{placeins} 

\usepackage{bm}
\usepackage{makecell}
\usepackage[flushleft]{threeparttable} 
\usepackage{graphicx}
\usepackage{amsmath}
\usepackage{amsfonts} 
\usepackage{commath} 
\usepackage{stackengine}
\usepackage[T1]{fontenc} 
\stackMath
\newcommand\tsup[2][2]{
 \def\useanchorwidth{T}%
  \ifnum#1>1%
    \stackon[-.5pt]{\tsup[\numexpr#1-1\relax]{#2}}{\scriptscriptstyle\sim}%
  \else%
    \stackon[.5pt]{#2}{\scriptscriptstyle\sim}%
  \fi%
}
\usepackage{tabularx, booktabs}
\usepackage{multirow}
\usepackage[flushleft]{threeparttable}
\usepackage{makecell}
\usepackage{adjustbox}
\usepackage{appendix}
\usepackage[figuresright]{rotating} 
\usepackage{hyperref}
\usepackage{cleveref}
\usepackage{times} 
\usepackage{authblk}
\usepackage[round]{natbib}

\newtheorem{Condition}{Condition}

\usepackage[plain,noend]{algorithm2e}

\usepackage{caption}
\usepackage{tikz-cd}
\usetikzlibrary{arrows}
\usepackage{tikz}
\usetikzlibrary{arrows.meta}

\tikzset{commutative diagrams/.cd, arrow style=tikz,diagrams={>= triangle 45}}

\makeatletter
\renewcommand{\algocf@captiontext}[2]{#1\algocf@typo. \AlCapFnt{}#2} 
\def\@algocf@capt@plain{top}
\renewcommand{\algocf@makecaption}[2]{%
  \addtolength{\hsize}{\algomargin}%
  \sbox\@tempboxa{\algocf@captiontext{#1}{#2}}%
  \ifdim\wd\@tempboxa >\hsize
    \hskip .5\algomargin%
    \parbox[t]{\hsize}{\algocf@captiontext{#1}{#2}}
  \else%
    \global\@minipagefalse%
    \hbox to\hsize{\box\@tempboxa}
  \fi%
  \addtolength{\hsize}{-\algomargin}%
}
\makeatother

\usepackage{amsmath,amsthm}
      \theoremstyle{plain}
      
      \newtheorem{theorem}{Theorem}

\usepackage[symbol]{footmisc}

\usepackage[pagewise]{lineno}

\begin{document}

\doublespacing

\title{Microbial correlation: a semi-parametric model for investigating microbial co-metabolism}



\author{Haoran Shi, Dan Cheng}
\affil{School of Mathematical and Statistical Sciences, Arizona State University
}

\author{Yue Wang}
\affil{Department of Bioinformatics and Informatics, Colorado School of Public Health, University of Colorado Anschutz Medical Campus
}

\maketitle

\begin{abstract}
The gut microbiome plays a crucial role in human health, yet the mechanisms underlying host-microbiome interactions remain unclear, limiting its translational potential. Recent microbiome multiomics studies, particularly paired microbiome-metabolome studies (PM2S), provide valuable insights into gut metabolism as a key mediator of these interactions. Our preliminary data reveal strong correlations among certain gut metabolites, suggesting shared metabolic pathways and microbial co-metabolism. However, these findings are confounded by various factors, underscoring the need for a more rigorous statistical approach.
Thus, we introduce microbial correlation, a novel metric that quantifies how two metabolites are co-regulated by the same gut microbes while accounting for confounders. 
Statistically, it is based on a partially linear model that isolates microbial-driven associations, and a consistent estimator is established based on semi-parametric theory. 
To improve efficiency, we develop a calibrated estimator with a parametric rate, maximizing the use of large external metagenomic datasets without paired metabolomic profiles. This calibrated estimator also enables efficient p-value calculation for identifying significant microbial co-metabolism signals. 
Through extensive numerical analysis, our method identifies important microbial co-metabolism patterns for healthy individuals, serving as a benchmark for future studies in diseased populations. 
\end{abstract}


\section{Introduction}
\label{sec:intro}

The gut microbiome has received significant attention in the past decade due to its observed associations with human health and disease. Studies demonstrate that gut microbes play critical roles in digestion, nutrient absorption, and immune system modulation \citep{jandhyala2015role, valdes2018role} as well as linked with colorectal cancer \citep{ahn2013human}, inflammatory bowel disease (IBD) \citep{franzosa2019gut} and 
diabetes \citep{zhao2018gut}. 
These findings underscore the potential of the gut microbiome for therapeutic and preventive applications, 
but the underlying mechanism behind the host-microbiome associations still remains largely unclear, limiting its translational impact. 
Recent microbiome multiomics studies—including genomics, transcriptomics, and metabolomics—provide new resources to advance our understanding of such mechanisms \citep{jiang2019microbiome, integrative2019integrative}. 
Among these studies, paired microbiome-metabolome studies (PM2S) are among the most promising, as they provide unique insights into gut metabolism, a potential mechanism by which the gut microbiome influences host health \citep{muller2022gut}.
Recent research has found that gut metabolites can influence host health in both beneficial and harmful ways.
For instance, butyrate, a short-chain fatty acid produced by gut bacteria, exerts anti-cancer effects by promoting apoptosis and maintaining intestinal barrier integrity, while deoxycholic acid, a secondary bile acid derived by the gut microbiome, has been linked to colorectal cancer progression by inducing DNA damage and inflammation \citep{zeng2019secondary}.

\begin{figure}
    \centering
    \begin{minipage}{0.48\textwidth}
    {\raggedright
    A \par}
        \includegraphics[trim={1cm 0cm 0cm 0cm},clip,width=1.09\linewidth]{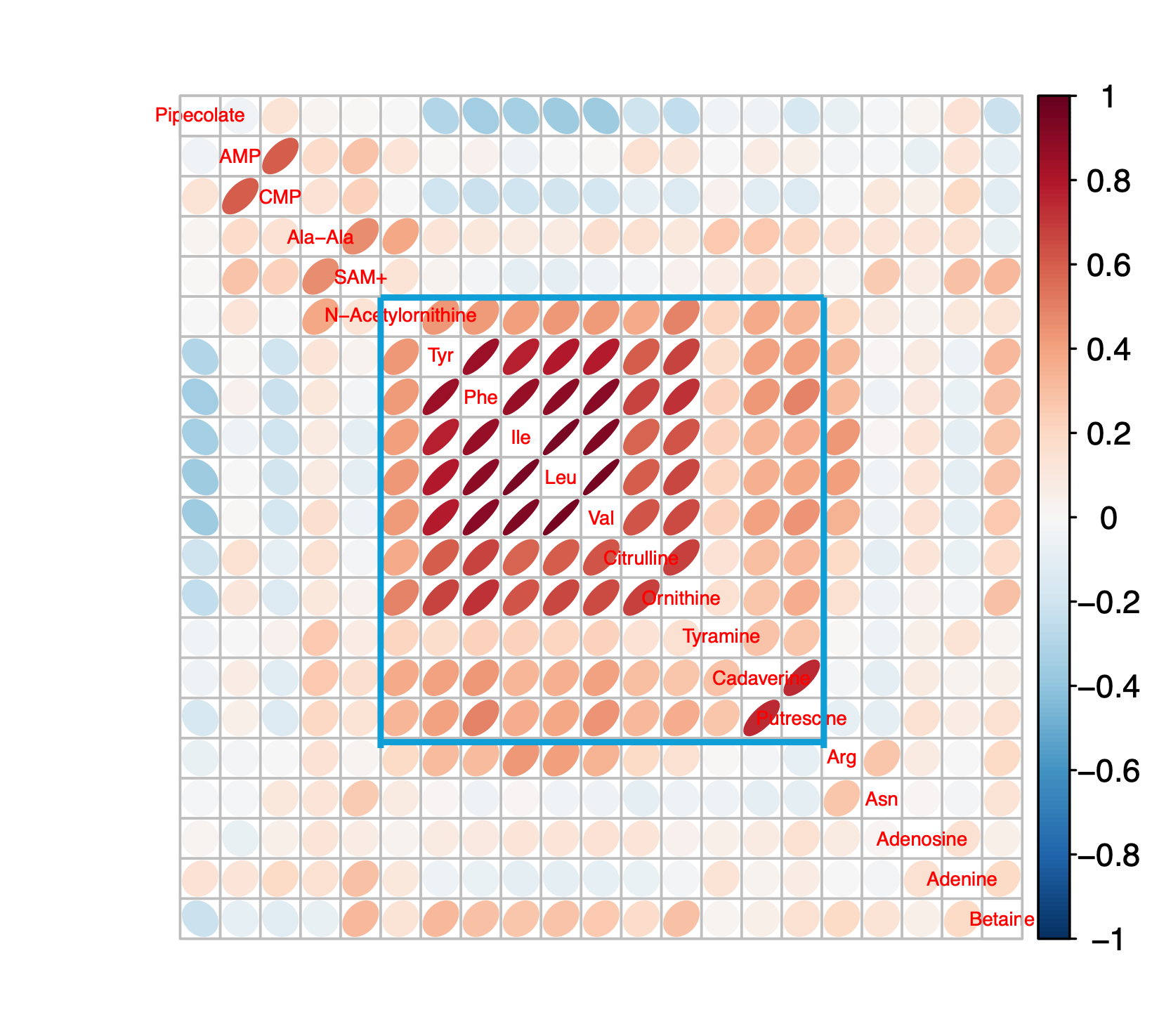}
    \end{minipage}\hfill
    \begin{minipage}{0.48\textwidth}
        {\raggedright
        B \par}
    \includegraphics[scale=0.6]{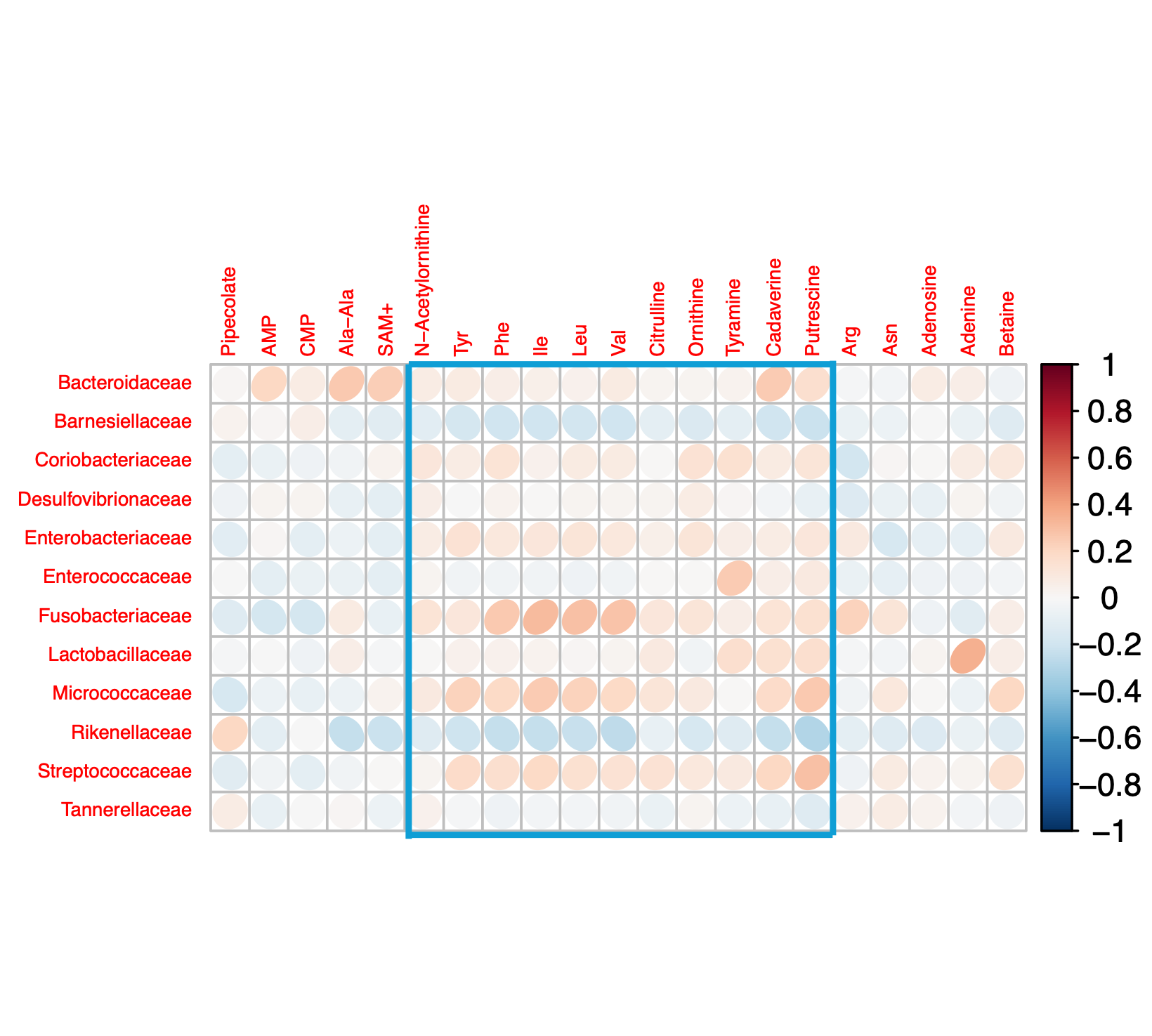}
    \end{minipage}\hfill
    \caption{A. Pairwise metabolome correlations with the names of metabolites on the diagonal. B. Microbiome-metabolome correlations with rows and columns representing microbes and metabolites, respectively. 
    In both figures, 
    red colors indicate positive correlations, while blue colors indicate negative correlations.}
    \label{fig:heatmap}
\end{figure}

Our motivating data was based on the 127 healthy individuals from the Yachida PM2S cohort \citep{yachida2019metagenomic}. 
Due to the excess zeros in the metagenomic data, we aggregated the species-level data into the microbial families and 
selected 12 microbial families present in at least 20\% of the subjects.  
More details of the Yachida cohort will be explained in Section \ref{sec:application}. 
Using the microbiome regression-based kernel association test (MiRKAT, \citealp{zhao2015testing}), we identified 21 gut metabolites significantly associated with the selected 12 microbial families.
Figure \ref{fig:heatmap}A shows a heatmap characterizing the Pearson correlation between the identified metabolites. 
We observed a major cluster consisting of 11 positively correlated metabolites. 
This may be because they originate from shared metabolic pathways and microbial co-metabolism. 
{Thus, we hypothesize that many gut metabolites interact with similar gut microbes in similar ways, thereby exhibiting strong correlations.} 
To have an initial test of this hypothesis, we further calculated the Pearson correlation between each microbe and each metabolite and presented the results in 
Fig. \ref{fig:heatmap}B. 
While these individual correlations were weak, which is a biological reality as gut metabolism involves the collaboration of many factors, 
we can still see that the 11 strongly correlated gut metabolites exhibit similar correlation patterns with the microbes. 
Specifically, they all showed positive correlations with Fusobacteriaceae and Micrococcaceae, while exhibited negative correlations with Rikenellaceae and Barnesiellaceae. 
While these preliminary data support our hypothesis regarding the existence of microbial co-metabolism, they are correlation-based, univariate and descriptive in nature, and thus, may be impacted by various confounding factors. 
For example, demographic factors such as age and sex, as well as lifestyle factors like diet and medication use, can significantly influence the gut microbiome and its metabolism \citep{yatsunenko2012human,kim2022sex}.
Additionally, since gut microbes frequently interact with each other, microbial confounders may also exist -- meaning a microbe may appear correlated with a metabolite simply because it is correlated with another microbe that directly produces that metabolite. 
{These limitations underscore the urgent need for a more statistically rigorous approach to studying microbial co-metabolism while accounting for various confounders. }



Our primary contribution is the development of microbial correlation, a metric that quantifies the degree to which two metabolites are co-regulated by the same set of gut microbes. 
Microbial correlation is based on partially linear models, combining a non-parametric component to account for potential non-linear effects of demographic and lifestyle confounders with a parametric component that enables the simultaneous analysis of multiple microbes.
As a result, unlike Pearson's correlations shown in Fig. \ref{fig:heatmap}, microbial correlation isolates the relationship between metabolites that is solely driven by gut microbes, effectively eliminating confounding effects. 
We construct a consistent estimator of the proposed microbial correlation with an established convergence rate of $o_p(n^{-1/4})$. 
To improve the efficiency, 
we further develop a calibrated estimator with a parametric convergence rate using sample-splitting and leveraging external metagenomic datasets.  
This calibration procedure is both feasible and meaningful. 
It is feasible because it relies only on external microbiome data without requiring paired metabolomic data, given the scarcity of paired microbiome-metabolome datasets. 
Importantly, it enables the integration of large-scale microbiome datasets, which would otherwise be unusable for studying microbial co-metabolism, thereby maximizing data utilization and integration. 
Furthermore, we establish the asymptotic normality of the calibrated estimator, facilitating efficient p-value calculations for hypothesis testing.  
To validate the proposed microbial correlation, along with its estimation and inference procedures, we conduct extensive simulation studies and real-world applications to study microbial co-metabolism patterns in the healthy population.

Throughout the paper, we use normal typeface to denote scalars, bold lowercase typeface to denote vectors, and uppercase typeface to denote matrices.  For any vector ${\bf v} \in \mathbb{R}^p$, we use $v_j$ to denote the $j$-{th} element of $\bf v$ for $j = 1, \ldots, p$. For any matrix ${M} \in \mathbb{R}^{n \times p}$, let $ {\bf m}_j$ denote the  $j$-{th} column of $M$ for $j = 1, \ldots, p$.
We use $m_{ij}$ or $(M)_{ij}$ to denote the $(i,j)$ entry of $M$ for $i = 1, \ldots, n$ and $j = 1, \ldots, p$. We use ${M}^\intercal$ to denote the transpose of the matrix ${M}$. We use ${M}'$ or ${\bf v}'$ to denote the derivative. 
Let $I(\mathcal{A})$ denote the indicator function of the event $\mathcal{A}$; i.e., $I(\mathcal{A}) = 1$ if $\mathcal{A}$ is true, and $I(\mathcal{A}) = 0$ otherwise. We use $A\bigotimes B$ to denote Kronecker product of two matrices $A$ and $B$. 
We use $\|\cdot\|$ to denote Euclidean norm. 
For ${M} \in \mathbb{R}^{n \times p}$, $\mbox{Vec}(M)$ is the $np \times 1$ column vector obtained by stacking the columns of the matrix $M$ on top of one another.

\section{Methodology}
\label{sec:meth}
\subsection{Partially linear model for analyzing microbial co-metabolism} \label{sec:set-up}
Suppose we observe independent and identically distributed data of $n$ subjects.
We use ${\bf x}_i \in \mathbb{R}^p$ to denote the microbial abundances of $p$ taxa, $\bm{z}_i \in \mathbb{R}^q$ for the confounding factors, and $y_i, w_i \in \mathbb{R}$ to represent two gut metabolites of interest. 
Given the compositional nature of microbiome data, we apply log-ratio transformations to each ${\bf x}_{i}$, such as 
the 
centered log-ratio transformation or the additive log-ratio transformation \citep{aitchison1982statistical}.



As discussed in Section \ref{sec:intro}, our goal is to quantify the correlation between $y_i$ and $w_i$ that is driven by the microbiome. 
To account for potential nonlinear confounding effects, 
we consider the following partial linear regression models
\begin{align}\label{1}
    \ y_{i}= f(\bm{z}_{i})+\boldsymbol{\beta}^\intercal {\bf x} _{i}+\epsilon_{i} ,\ 
w_{i}= g(\bm{z}_{i})+\boldsymbol{\gamma}^\intercal{\bf x}_{i}+\delta_{i}, 
\end{align}
  where $\epsilon_i$ and $\delta_i$ are independent random variables, assumed 
independent of $\bm{z}_i$ and ${\bf x}_i$. 
We also assume 
  $\mathbb{E}(\epsilon_i) = 0, \mathbb{E}(\delta_i) = 0$, $\mathbb{E}(\epsilon_i^2)<\infty$ and $\mathbb{E}(\delta_i^2)<\infty$,
  $f(\cdot), g(\cdot): \mathbb{R}^p \rightarrow \mathbb{R}$ are two unknown smooth functions, 
and $\boldsymbol{\beta}, \boldsymbol{\gamma} \in \mathbb{R}^p$ characterize the relationship between the microbiome and the two metabolites. 
  
  

Based on \eqref{1}, we next develop the microbial correlation between $y_i$ and $w_i$.  
We first rewrite \eqref{1} as 
\begin{align}\label{1:rev}
y_{i}-h_y(\bm{z}_i)=\boldsymbol{\beta}^\intercal ({\bf x}_i- h_x(\bm{z}_i))+ \epsilon_i,~~~  w_{i}- h_w(\bm{z}_i)=\boldsymbol{\gamma}^\intercal ({\bf x}_i- h_x(\bm{z}_i))+ \delta_i, 
\end{align}
where $h_x(\bm{z}_i) = \mathbb{E}({\bf x}_i \mid \bm{z}_i)$, $h_w(\bm{z}_i) = \mathbb{E}(w_i \mid \bm{z}_i)$ and $h_y(\bm{z}_i) = \mathbb{E}(y_i \mid \bm{z}_i)$.
Based on \eqref{1:rev}, we define the microbial correlation between $y_i$ and $w_i$ as
\begin{align}\label{microbial-correlation}
R = \mbox{Cor}\left( \boldsymbol{\beta}^\intercal ({\bf x}_i- h_x(\bm{z}_i)), \boldsymbol{\gamma}^\intercal ({\bf x}_i- h_x(\bm{z}_i)) \right).
\end{align}

Our microbial correlation $R$ is defined to isolate the relationship between metabolites that is solely driven by gut microbes, effectively eliminating confounding effects.
Specifically, unlike Pearson’s correlation which is a general correlation measure, 
our microbial correlation focuses entirely on metabolite-wise relationships driven by the gut microbiome, 
offering biological insights into microbial co-metabolism. 



Our microbial correlation is related to the several existing work on genetic correlation in the context of genome-wide association studies (GWAS), 
where the goal is to identify genetically related traits.  
For example, assuming linear models between the traits $y_i, w_i$ and the genetic data ${\bf x}_i$, i.e., 
$y_i = {\bf x}_i^\intercal \boldsymbol{\beta} + \epsilon_i$ and 
$w_i = {\bf x}_i^\intercal \boldsymbol{\gamma} + \delta_i$, 
\cite{guo2019optimal} define the genetic relatedness as $R_1 = { \bm {\beta}^\intercal \bm {\gamma} }/{\sqrt{ \bm {\beta}^\intercal \bm {\beta}  \bm {\gamma}^\intercal \bm {\gamma}} }$,  
i.e., the cosine value of the Euclidean angle between $\boldsymbol{\beta}$ and $\boldsymbol{\gamma}$. 
One limitation of $R_1$ arises from its fully deterministic nature, failing to account for the variations in the genetic data. Also, it
cannot be interpreted as a correlation coefficient. 
To address this limitation,
\cite{wang2022estimation} and \cite{ma2022statistical} define an alternative genetic correlation measure ${R}_2 = \mbox{Cor}(\boldsymbol{\beta}^\intercal {\bf x}_{i}, \boldsymbol{\gamma}^\intercal {\bf x}_{i})$ 
by incorporating variations from the genetic data. 
However, neither $R_1$ nor $R_2$ account for potential confounders, making them less suitable for analyzing microbial co-metabolism, as many factors beyond the gut microbiome can influence gut metabolism.
Our definition of the microbial correlation addresses this limitation by accounting for nonlinear effects of confounders through the partial linear models in \eqref{1}. 

We construct a simple example to further illustrate the key differences among $R, R_1$ and $R_2$. 
Consider a simplified version of model \eqref{1:rev} with 
\(
{\bf x}_i = B\bm{z}_i + \bm{\xi}_i,
\)
where $\bm{\xi}_i$ is independent of $\bm{z}_i$ with $E[\bm{\xi}_i]=0$, $\mbox{Cov}(\bm{z}_i) = I_q$, and $B^\intercal B = I_q$ with $B \in \mathbb{R}^{p \times q}$. 
Under this model, some algebra leads to the following explicit forms of $R, R_1$, and $R_2$: 
\begin{align*}
R = &~ \frac{\boldsymbol{\beta} ^\intercal \Phi \boldsymbol{\gamma}}{ \sqrt{ \boldsymbol{\beta} ^\intercal \Phi \boldsymbol{\beta} }\sqrt{ \boldsymbol{\gamma} ^\intercal \Phi \boldsymbol{\gamma} }}, ~~~ R_1 = \frac{ \bm {\beta}^\intercal \bm {\gamma} }{\sqrt{ \bm {\beta}^\intercal \bm {\beta} }\sqrt{ \bm {\gamma}^\intercal \bm {\gamma}} }, ~~~ 
R_2 = \frac{\boldsymbol{\beta} ^\intercal (\Phi + \Sigma)\boldsymbol{\gamma}}{ \sqrt{\boldsymbol{\beta} ^\intercal (\Phi + \Sigma) \boldsymbol{\beta} }\sqrt{\boldsymbol{\gamma} ^\intercal (\Phi + \Sigma) \boldsymbol{\gamma} }}, 
\end{align*}
where $\Phi = \mbox{Cov}(\bm{\xi}_i)$ and $\Sigma = BB^\intercal$. 
Consider the first scenario where $\boldsymbol{\beta}^\intercal \boldsymbol{\gamma} = 0$ and ${\Phi} = I_p$. 
In this case, one can derive that $R = R_1 = 0$ but $R_2 \neq 0$ due to the addition term $\Sigma$ in the explicit form of $R_2$. 
Consider another scenario where  $\boldsymbol{\beta}^\intercal \boldsymbol{\gamma} = 0$, and 
$\Phi = \widetilde{B}\widetilde{B}^\intercal$, where $\widetilde{B} \in \mathbb{R}^{p \times (p-q)}$ with  $\widetilde{B}^\intercal \widetilde{B} = I_{p-q}$ and $\widetilde{B}^\intercal B = 0$. 
In this case, since $\Phi + \Sigma = I_p$, 
one can show that
$R_1 = R_2 = 0$ but $R \neq 0$. 
From these examples, we can see that $R, R_1$, and $R_2$ may yield substantially different values depending on the specific scenario. 
This also highlights the importance of accounting for potential confounding variables for studying microbial co-metabolism.  

\subsection{Estimation}\label{sec:est}

In this section, we develop a consistent estimator of the proposed microbial correlation $R$. 
Recall from \eqref{1:rev} that 
\(
R = \mbox{Cor}\left( \boldsymbol{\beta}^\intercal ({\bf x}_i- h_x(\bm{z}_i)), \boldsymbol{\gamma}^\intercal ({\bf x}_i- h_x(\bm{z}_i)) \right). 
\)
Some algebra leads to 
\[
R = \frac{\boldsymbol{\beta} ^\intercal \Phi \boldsymbol{\gamma}}{ \sqrt{ \boldsymbol{\beta} ^\intercal \Phi \boldsymbol{\beta} }\sqrt{ \boldsymbol{\gamma} ^\intercal \Phi \boldsymbol{\gamma} }}, 
\]
where $\Phi = \mbox{Cov}({\bf x}_i - h_x(\bm{z}_i))
= \mathbb{E}[({\bf x}_i - h_x(\bm{z}_i)) ({\bf x}_i - h_x(\bm{z}_i))^\intercal]$.
Based on this alternative from, to estimate $R$, we need to estimate
$\boldsymbol{\beta}$, $\boldsymbol{\gamma}$, $h_x(\cdot)$, and $\Phi$.  

First, we construct the Nadaraya–Watson kernel estimator of $h_x(\cdot)$ \citep{watson1964smooth,nadaraya1964estimating}.
Given a kernel function
\(k: \mathbb{R} \to \mathbb{R}\),
we define
\begin{align}\label{kernel}
K(\bm{z}_i) = \prod_{j=1}^q k(z_{ij}),
\quad K_{ij} = K\left( \frac{\bm{z}_i - \bm{z}_j}{a} \right)
\end{align}
where \(z_{ij}\) is the \(j\)-th element of \(\bm{z}_i\), and \(a\) is a positive constant called the bandwidth.  
We then estimate $h_x(\bm{z}_i)$ with 
\(
\widehat{h}_x(\bm{z}_{i}) = {\widehat{l}_i}^{-1}{\sum_{j=1}^n {\bf x}_i K_{ij}}
\)
with 
\(
\widehat{l}_i = (na^q)^{-1} \sum_{j=1}^n K_{ij}. 
\)
Similarly, we estimate $h_y(\bm{z}_i)$ and $h_w(\bm{z}_i)$
with 
\(
\widehat{h}_y(\bm{z}_{i}) = {\widehat{l}_i}^{-1}{\sum_{j=1}^n y_i K_{ij}}
\)
and 
\(
\widehat{h}_w(\bm{z}_{i}) = {\widehat{l}_i}^{-1}{\sum_{j=1}^n w_i K_{ij}}, 
\)
respectively.



Next, we estimate $\boldsymbol{\beta}$ and $\boldsymbol{\gamma}$ based on model \eqref{1:rev}. 
Specifically, we plug-in $\widehat{h}_x(\cdot), \widehat{h}_y(\cdot), \widehat{h}_w(\cdot)$ into \eqref{1:rev} and get 
\begin{align}\label{4}
y_{i}- \widehat{h}_y(\bm{z}_i) =\boldsymbol{\beta}^\intercal ({\bf x}_i-\widehat{h}_x(\bm{z}_i))+ \epsilon_i, ~~~  w_{i}- \widehat{h}_w(\bm{z}_i) =\boldsymbol{\gamma}^\intercal ({\bf x}_i-\widehat{h}_x(\bm{z}_i))+ \delta_i.
\end{align}




 
Following \cite{1988Root}, we estimate $\Phi$ using a truncated covariance estimator:
\begin{align}\label{h:est}
    \widehat{\Phi}=n^{-1}\sum_{i=1}^{n} \{{\bf x}_i-\widehat{h}_x(\bm{z}_{i})\}\{{\bf x}_i-\widehat{h}_x(\bm{z}_{i})\}^\intercal {I}(\widehat{l}_i > b),
\end{align}
where $b$ is a truncation parameter. 
Then, we estimate \(\boldsymbol{\beta}\) and \(\boldsymbol{\gamma}\) using the truncated ordinary least squares:
\begin{align}\label{beta}
    \widehat{\boldsymbol{\beta}} = &~ 
    n^{-1}\widehat{\Phi}^{-1}
    \sum_{i=1}^n \left\{({\bf x}_i-\widehat{h}_x(\bm{z}_{i}))(y_i-\widehat{h}_y(\bm{z}_{i})){I}(\widehat{l}_i > b)\right\} ,
\nonumber \\
    \widehat{\boldsymbol{\gamma}} = &~ 
    n^{-1}\widehat{\Phi}^{-1}  \sum_{i=1}^n \left\{({\bf x}_i-\widehat{h}_x(\bm{z}_{i}))(w_i-\widehat{h}_w(\bm{z}_{i})){I}(\widehat{l}_i > b)\right\} ,
\end{align}
where ${I}(\cdot)$ is an indicator function. 
In the truncated estimators above, only subjects with a sufficiently large value of $l_i$ are included. 

Finally, we estimate the microbial correlation $R$ with a plug-in estimator: 
\begin{align}\label{Rhat}
\widehat{R} = \frac{ \widehat{\boldsymbol{\beta}} ^\intercal \widehat{\Phi} \widehat{\boldsymbol{\gamma}} }{ \sqrt{ \widehat{\boldsymbol{\beta}} ^\intercal \widehat{\Phi}\widehat{\boldsymbol{\beta}} }\sqrt{ \widehat{\boldsymbol{\gamma}} ^\intercal \widehat{\Phi}\widehat{\boldsymbol{\gamma}} }}. 
\end{align}

We next study the consistency of $\widehat{R}$. We introduce the following conditions. 

\begin{Condition} \label{3.1}
Each element of $\bm{z}_i$ follows a sub-exponential distribution, that is, $\operatorname{P}(|z_{ij}|>z)\leq e^{-C_{j}z}$ for some constant $C_{j}>0$ for $i=1, \ldots, n$ and $j = 1, \ldots, q$. 
\end{Condition}



\begin{Condition} \label{3.2}
For the kernel function $k(u)$ defined in \eqref{kernel}, there exists a positive integer $m > q/2$ such that  
for each $s=1, \dots, m-1$, we have 
    \begin{equation*}
\int_{\mathcal{R}} k(u) \, du = 1, ~~ \int_{\mathcal{R}} u^s k(u) \, du = 0, ~~~ k(u) = O\left((1 + |u|^{m+1+\epsilon})^{-1}\right) \mbox{  as  } |u| \rightarrow \infty,
\end{equation*}
for some $\epsilon >0$. 
\end{Condition}
\begin{Condition}\label{3.3}
     As $n \to \infty$, $n a^{2q} b^4 \to \infty$, $n a^{4m}b^{-4}\to 0$ , $a^{m}b^{-2} \to 0$, and $\ b \to 0$, 
     where 
     $a$ is the kernel bandwidth defined in \eqref{kernel}, $b$ is the cutoff used in \eqref{h:est} and  \eqref{beta}, $m$ is defined in Condition \ref{3.2}, and $q$ is the number of covariates. 
\end{Condition}

Condition \ref{3.2} requires
the integral of \( k(u) \) over all \( u \) is 1, ensuring that the kernel properly weights the data it smooths. 
It also requires the first $m-1$ ``moments" to be zero, ensuring that the kernel is centered at zero and provides unbiased smoothing. 
The decay condition ensures that the tail of the kernel function diminishes quickly as \( |u| \) increases, which helps limit the effect of distant data points and reduces the variance of the estimates produced by the kernel. 
With $m=2$, a popular kernel choice that satisfies Condition \ref{3.2} is the Gaussian kernel $k(u)=(2\pi)^{-1/2} \exp \{-{u^2}/{2}\}$ 
because $\int k(u) du = 1$, $\int u k(u) du = 0$. 
Also, the exponential function decays faster than any polynomial functions, indicating the Gaussian kernel is $O\left((1 + |u|^{3+\epsilon})^{-1}\right)$ for any $\epsilon > 0$. 
However, in this case, we require $q \leq 3$, indicating that we can adjust at most 3 covariates. 
This limitation of using Gaussian kernels for partially linear models is also discussed in 
\cite{1988Root}.
Nonetheless, 
one can construct
a higher-order kernel satisfying Condition \ref{3.2} with an arbitrarily large $m$ by a weighted sum of Gaussian kernels; the readers are referred to \cite{rao2014nonparametric} for more details. 
Moreover, such higher-order kernels are supported in existing {\it R} packages for fitting partially linear models, such as {\tt np} \citep{np}. 
In addition, when adjusting for many covariates (i.e., $q$ is large), one can instead consider the generalized additive model as a special case of \eqref{1}, i.e., 
\(
y_i = \sum_{j=1}^q h_{y,j}(z_{ij}) + \boldsymbol{\beta}^\intercal {\bf x}_i + \epsilon_i, ~~ w_i = \sum_{j=1}^q h_{w,j}(z_{ij}) + \boldsymbol{\gamma}^\intercal {\bf x}_i + \delta_i. 
\)
In this case, since each covariate is modeled separately in an additive form, Gaussian kernels are still valid for estimating each $h_{y,j}(\cdot)$ and $h_{w,j}(\cdot)$.

Condition \ref{3.3} requires both $a$ and $b$ to converge to 0, while $b$ can converge at an arbitrarily slow rate. 
When using Gaussian kernels, i.e., $m=2$, $a$ is required to converge to 0 at a faster rate than $b$, 
because $a^m b^{-2}$ is required to converge to 0. 
In particular, when \( q = 2 \) and $m=2$,  Condition \ref{3.3} simplifies to \( b \rightarrow 0 \), \( a/b \rightarrow 0 \), \(n^{-1/4}a^2b^{-1} \to 0\) and \( n^{1/4}ab \rightarrow \infty \) as \( n \rightarrow \infty \). These conditions can be satisfied if we set \( a = n^{-1/6} \) and \( b = n^{-1/12 + \varepsilon} \) for any \( \varepsilon \in (0, 1/12) \).





Our first result characterizes the consistency of $\widehat{R}$ given in \eqref{Rhat}. 
\begin{theorem}\label{theorem1} 
  Suppose Conditions \ref{3.1}--\ref{3.3} hold. As $n \rightarrow \infty$, we have
    \begin{align*}
      \widehat{R} - R = o_p(n^{-1/4}). 
    \end{align*}
\end{theorem}

Despite the consistency of $\widehat{R}$, it has two limitations. 
First, its convergence rate is slower than \( n^{-1/2} \) due to the slow convergence rate of $\widehat{\Phi}$ ({see Lemma 1.1 in the supplement}), 
making it less efficient in real applications. 
Second, $\widehat{R}$ does not have a tractable asymptotic distribution due to the 
complex dependencies among $\widehat{\boldsymbol{\beta}}, \widehat{\boldsymbol{\gamma}}$ and $\widehat{\Phi}$ that arise from the kernel regression. 
This further limits the use of the proposed microbial correlation, as its theoretical confidence interval cannot be constructed.  


\subsection{Inference} \label{sec:inf}
We address these issues by developing a calibrated estimation
with an improved convergence rate to facilitate efficient hypothesis testing.  
Specifically, we randomly divide the sample into two parts, separately used for estimating $\boldsymbol{\beta}$ and $\boldsymbol{\gamma}$ according to \eqref{beta} with the bandwidth $a$ and the cutoff $b$ that satisfy Condition \ref{3.3}, and the resulting estimators are denoted by 
$\widehat{\boldsymbol{\beta}}_{ss}$ and $\widehat{\boldsymbol{\gamma}}_{ss}$, respectively. 
We then use the external microbiome data with sample size $N$
to obtain an estimate of \( \Phi \) according to \eqref{h:est}, denoted by $\widehat{\Phi}_{ss}$; here, $a_{ss}$ and $b_{ss}$ are related to $N$, which 
satisfy Condition \ref{cond:aPhi:bPhi} that will be introduced later. 
This approach ensures the mutual independence among $\widehat{\Phi}_{ss}$,  \( \widehat{\boldsymbol{\beta}}_{ss} \) and \( \widehat{\boldsymbol{\gamma}}_{ss} \). 
Finally, we obtain a calibrated estimator of $R$: 
\begin{align}\label{Rhat:new}
\widehat{R}_{ss} = \frac{ \widehat{\boldsymbol{\beta}}_{ss} ^\intercal \widehat{\Phi}_{ss} \widehat{\boldsymbol{\gamma}}_{ss} }{ \sqrt{ \widehat{\boldsymbol{\beta}}_{ss} ^\intercal \widehat{\Phi}_{ss} \widehat{\boldsymbol{\beta}}_{ss} }\sqrt{ \widehat{\boldsymbol{\gamma}}_{ss} ^\intercal \widehat{\Phi}_{ss} \widehat{\boldsymbol{\gamma}}_{ss} }}
\end{align}

From a practical perspective, we do not require the external microbiome data to have paired metabolomic profiles. 
This is because the calculation of $\widehat{\Phi}_{ss}$ only relies on the metagenomic data ${\bf x}$ and the covariates $\bm{z}$ without using the metabolite information $y$ and $w$. 
This provides substantial flexibility for selecting external metagenomic data sets for estimating $\Phi$, as many large-scale metagenomics data do not have paired 
metabolomics data primarily due to the maturity of metagenomic sequencing technologies and the scarcity of paired microbiome-metabolome studies. 
This approach also maximizes the use of existing microbiome data for analyzing microbial metabolism. 
In our real application, we combine over 9,000 samples from 93 microbiome studies as the external data set to analyze microbial co-metabolism in healthy individuals. 

Next, we will show the $\sqrt{n}-$consistency and asymptotic normality of the calibrated estimator $\widehat{R}_{ss}$.
We require the following additional condition regarding $N, a_{ss}$ and $b_{ss}$, which is parallel to Condition \ref{3.3}. 

\begin{Condition}\label{cond:aPhi:bPhi}
As $n \rightarrow \infty$ and $N \rightarrow \infty$, we have $b_{ss} {n} \log n  \rightarrow 0$, $a_{ss}^{m}b^{-2}_{ss} \to 0$, $ N^{-1/2}  a_{ss}^{-q}  b_{ss}^{-2} \rightarrow 0$, $N a_{ss}^{4m}b_{ss}^{-4}\to 0$, 
where $m$ is the order of the kernel specified in Condition \ref{3.2}, and $q$ is the number of confounding factors. 
\end{Condition}

Condition \ref{cond:aPhi:bPhi} implies a relationship between the PM2S sample size $n$ and the external sample size $N$. 
To understand this, we consider a specific case 
where $n = N^{\alpha_1}, a_{ss} = N^{-\alpha_2}$ and $b_{ss} = N^{-\alpha_3}$ with $0 < \alpha_1 < 1, \alpha_2 > 0$ and $\alpha_3 > 0$. 
Thus, Condition \ref{cond:aPhi:bPhi} requires $1/(4m)<\alpha_{2}<\operatorname{min}(1/(2q),1/2(q+2))$ , $\alpha_{3}<(1-2q\alpha_{2})/4$, and $\alpha_1 < \alpha_3$. 
Since the kernel order $m$ can be selected arbitrarily large in real applications, we can derive $\alpha_1 < \alpha_3 < 1/4$ as $m \rightarrow \infty$. 
Thus, a necessary condition for Condition \ref{cond:aPhi:bPhi} to hold is that $n = o(N^{1/4})$. 
While this requires the external data set to have a large sample size in theory, our simulation studies in Section \ref{sec:simulation} show that with $N = 10n$, we have good estimation accuracy, well-controlled type-I error rates, and decent statistical power.

We have the next result to characterize the asymptotic distribution of the calibrated estimator $\widehat{R}_{ss}$ in \eqref{Rhat:new}. 

\begin{theorem}\label{theorem2}
Suppose Conditions \ref{3.1}-\ref{cond:aPhi:bPhi} hold. As $n, N \rightarrow \infty$, we have 
\[
 \sqrt{n}\frac{(\widehat{R}_{ss}-R)}{{\sigma_{\widehat{R}_{ss}}}} \stackrel{d}{\rightarrow} N(0,1), 
\]
where 
\begin{align*}
    \sigma^2_{\widehat{R}_{ss}}&=\frac{1}{\boldsymbol{\gamma}^\intercal \Phi \boldsymbol{\gamma} \boldsymbol{\beta}^\intercal \Phi \boldsymbol{\beta}} \Sigma_{11}
     +\frac{(\boldsymbol{\beta}^\intercal \Phi \boldsymbol{\gamma})^2}{4(\boldsymbol{\beta}^\intercal\Phi\boldsymbol{\beta})^3(\boldsymbol{\gamma}^\intercal\Phi\boldsymbol{\gamma})}\Sigma_{22}
    +\frac{(\boldsymbol{\beta}^\intercal \Phi \boldsymbol{\gamma})^2}{4(\boldsymbol{\gamma}^\intercal\Phi\boldsymbol{\gamma})^3(\boldsymbol{\beta}^\intercal\Phi\boldsymbol{\beta})}\Sigma_{33}\\
    &\quad -\frac{\boldsymbol{\beta}^\intercal\Phi \boldsymbol{\gamma}}{(\boldsymbol{\beta}^\intercal \Phi \boldsymbol{\beta})^2(\boldsymbol{\gamma}^\intercal \Phi \boldsymbol{\gamma})}\Sigma_{12}
    -\frac{\boldsymbol{\beta}^\intercal\Phi \boldsymbol{\gamma}}{(\boldsymbol{\beta}^\intercal \Phi \boldsymbol{\beta})(\boldsymbol{\gamma}^\intercal \Phi \boldsymbol{\gamma})^2}\Sigma_{13}+\frac{(\boldsymbol{\beta}^\intercal\Phi \boldsymbol{\gamma})^2}{2(\boldsymbol{\beta}^\intercal \Phi \boldsymbol{\beta})^2(\boldsymbol{\gamma}^\intercal \Phi \boldsymbol{\gamma})^2}\Sigma_{23}.
\end{align*}
Here $\Sigma_{ij}$ are functions of $\boldsymbol{\beta}$, $\boldsymbol{\gamma}$, $\Phi$, $\sigma^2_{\epsilon}$ and $\sigma^2_{\delta}$ with specific forms given in eq. (5) in the supplement. 
\end{theorem}





Theorem \ref{theorem2} provides the theoretical basis for testing the hypothesis \( H_0: |R| \leq R_{0} \) against \( H_1: |R| > R_{0} \) for some $R_0 \in [0,1)$ to identify strong signals of microbial co-metabolism. 
Specifically, we construct our test statistics as
\begin{align}\label{test:R0}
T_+ = \frac{ \sqrt{n} (|\widehat{R}_{ss}| - R_0)_{+}}{\widehat{\sigma}_{\widehat{R}_{ss}}}, ~~ T_- = \frac{ \sqrt{n} (|\widehat{R}_{ss}| - R_0)_{-}}{\widehat{\sigma}_{\widehat{R}_{ss}}}
\end{align}
where 
$\widehat{\sigma}^2_{\widehat{R}_{ss}}$ is a consistent estimator of ${\sigma}^2_{\widehat{R}_{ss}}$ that will be discussed later, 
$a_+ = \max(a, 0)$ and $a_- = \min(a, 0)$.
Then, the $p$-value for testing \( H_0: |R| \leq R_{0} \) is given by 
\begin{align}\label{pval:R0}
p = 2\mbox{Pr}(Z \geq \max(T_+, T_-)),
\end{align}
where $Z \sim N(0,1)$. 
Notably, when $R_0 = 0$, the $p$-value simplifies to $p = 2\mbox{pr}(Z \geq |T|)$, where 
\[
T = \frac{ \sqrt{n} |\widehat{R}_{ss}|}{\widehat{\sigma}_{\widehat{R}_{ss}}}. 
\]

To estimate ${\sigma}^2_{\widehat{R}_{ss}}$, which is a function of $\boldsymbol{\beta}, \boldsymbol{\gamma}, {\Phi}, \sigma_\epsilon^2$ and $\sigma_\delta^2$, 
we first estimate $\boldsymbol{\beta}, \boldsymbol{\gamma}, {\Phi}$ with \( \widehat{\boldsymbol{\beta}} \), \( \widehat{\boldsymbol{\gamma}} \), and \( \widehat{\Phi} \), respectively, given in \eqref{beta} and \eqref{h:est}. 
To estimate $\sigma_\epsilon^2$, 
we first obtain the residuals from \eqref{4} according to $\widehat{\epsilon}_i = y_i-\widehat{h}_y(\bm{z}_i) - \widehat{\boldsymbol{\beta}}^\intercal\{{\bf x}_i- \widehat{h}_x(\bm{z}_i) \}$ for $i = 1, \ldots, n$. 
We then use the mean squared error of $\widehat{\epsilon}_i$ to estimate $\sigma^2_{\epsilon}$. The estimate of $\sigma^2_{\delta}$ is obtained in the same way.

\section{Simulation Studies}\label{sec:simulation}


In this section, we conducted simulation studies to evaluate the performance of our proposed method under finite sample sizes. 
We first generated $q$-dimensional covariates $\bm{z}_i$ from a multivariate normal distribution with mean ${\bm{0}}$ and covariance ${I}_q$. 
We then generated the microbiome data ${\bf x}_i$ according to $x_{ij} = h_{x}(\bm{z}_{i}) + {\tau}_{ij}$ for $j = 1, \ldots, p$, where 
$\bm{\tau}_i = (\tau_{i1}, \ldots, \tau_{ip})^\intercal$ was generated from a multivariate normal distribution with mean 0 and covariance $\Phi$. 
For simplicity, we used the same $h_{x}(\cdot)$ for all components of ${\bf x}_i$, which will be specified later. 
The error terms $\epsilon_i$ and $\delta_i$ were simulated from independent standard normal distributions. 
Finally, the outcomes $y_i$ and $w_i$ were generated according to model \eqref{1}, where $y_i = f(\bm{z}_i) +\boldsymbol{\beta}^\intercal {\bf x}_i + \epsilon_i$ and $w_i = g(\bm{z}_i) +\boldsymbol{\gamma}^\intercal {\bf x}_i + \delta_i$, where
$f(\cdot)$ and $g(\cdot)$ will be specified below. 
We also generated an external data set $\{{\bf x}_i, \bm{z}_i\}$ for $i = 1, \ldots, N$ in the same way, where $N=10n$. 
Note that the external data is not required to have the paired metabolome data. 



We considered $n = 100, 300, 500$, $p=6$, and $q=2$. 
The covariance matrix of $\bm{\tau}_i, i.e., $ $\Phi$,  was generated in two scenarios: In the first scenario, we set $\Phi = I_p$, while in the second scenario, 
$\Phi = DD^\intercal$, where $D$ is an upper triangular matrix
with diagonal entries being 1 and off-diagonal elements being 0.5. 
We considered three scenarios of $h_x(\cdot), g(\cdot)$, and $f(\cdot)$:
\begin{align*}
\mbox{Linear:} &~~ h_{x}(\bm{z}_{i}) = f(\bm{z}_i) = g(\bm{z}_i) = z_{i1} + z_{i2}, \\
\mbox{Exponential:} &~~ h_{x}(\bm{z}_{i}) = f(\bm{z}_i) = g(\bm{z}_i) = \exp\{\frac{-(z_{i1} + z_{i2})^2}{2}\}, \\
\mbox{Triangular:} &~~ h_{x}(\bm{z}_{i}) = f(\bm{z}_i) = \operatorname{sin}(z_{i1} + z_{i2}); ~~ g(\bm{z}_i) = \operatorname{cos}(z_{i1} + z_{i2}).
\end{align*}
As $R$ depends on $\boldsymbol{\beta},\boldsymbol{\gamma},$ and $\Phi$, 
we generated $\boldsymbol{\beta}$ and ${\boldsymbol{\gamma}}$ in a way that can control the value of $R$ for better illustration. 
For a pre-specified value of $R = R_0$, 
in the first scenario where $\Phi$ is an identity matrix,
we set 
\begin{align*}
&\boldsymbol{\beta} = (\sqrt{3}/3, \sqrt{3}/3, \sqrt{3}/3, 0, 0, 0)^\intercal,  \\
&\boldsymbol{\gamma} = \left(R_0/\sqrt{3}, R_0/\sqrt{3}, R_0/\sqrt{3}, \sqrt{(1-R_0^2)/3}, \sqrt{(1-R_0^2)/3}, \sqrt{(1-R_0^2)/3}\right)^\intercal.
\end{align*}
In the second scenario where $\Phi = DD^\intercal$, 
we adjusted $\boldsymbol{\gamma}$ and $\boldsymbol{\beta}$ using $D^{-1}$:
\begin{align*}
&\boldsymbol{\beta} = D^{-1}(\sqrt{3}/3, \sqrt{3}/3, \sqrt{3}/3, 0, 0, 0)^\intercal ,  \\
&\boldsymbol{\gamma} = D^{-1} \left(R_0/\sqrt{3}, R_0/\sqrt{3}, R_0/\sqrt{3}, \sqrt{(1-R_0^2)/3}, \sqrt{(1-R_0^2)/3}, \sqrt{(1-R_0^2)/3}\right)^\intercal .
\end{align*}
In both scenarios, one can verify that the true microbial correlation defined in \eqref{microbial-correlation} equals $R_0$.




\begin{figure}[htbp]
    \centering
    \begin{minipage}{0.48\textwidth}
    {\raggedright
A \par}
        \centering
        \includegraphics[scale=0.6]{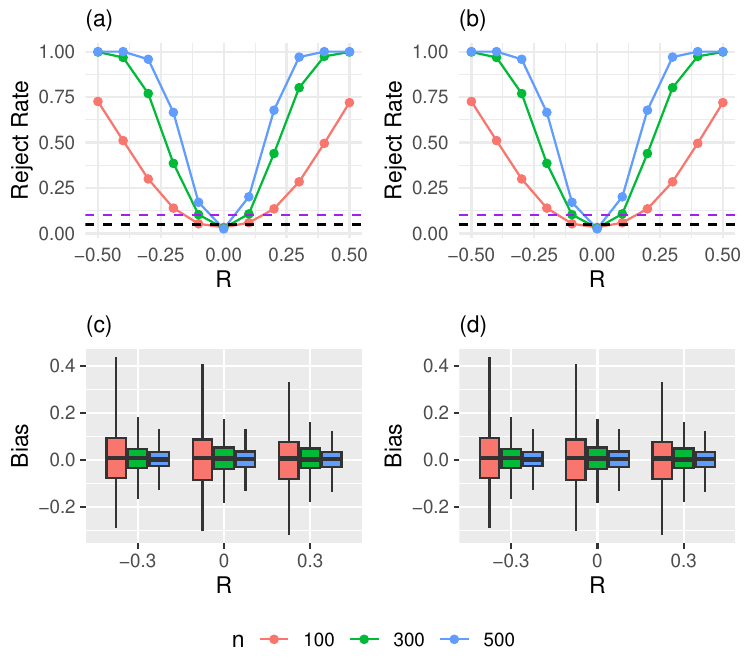}
    \end{minipage}\hfill
    \begin{minipage}{0.48\textwidth}
        {\raggedright
B \par}
        \centering
        \includegraphics[scale=0.6]{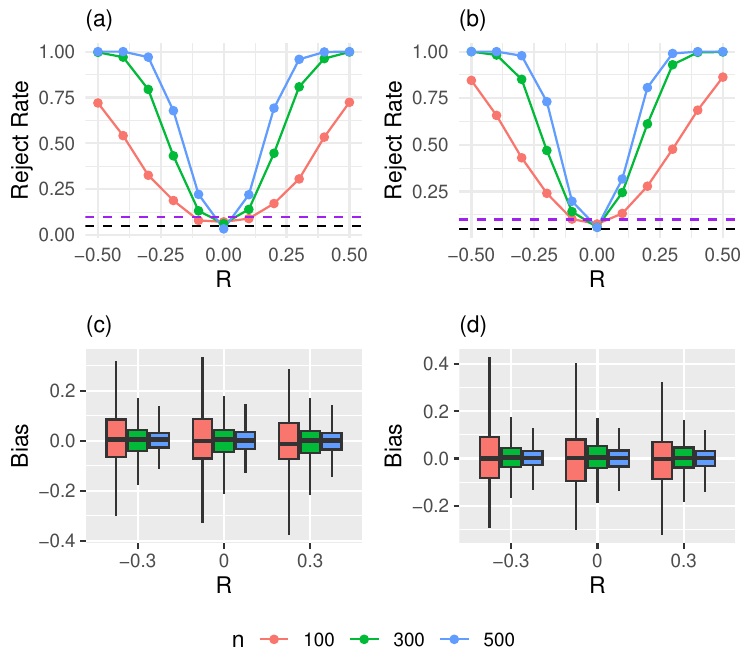}

    \end{minipage}

    \vspace{0.5cm}

    \begin{minipage}{0.48\textwidth}
        {\raggedright
C \par}
        \centering
        \includegraphics[scale=0.6]{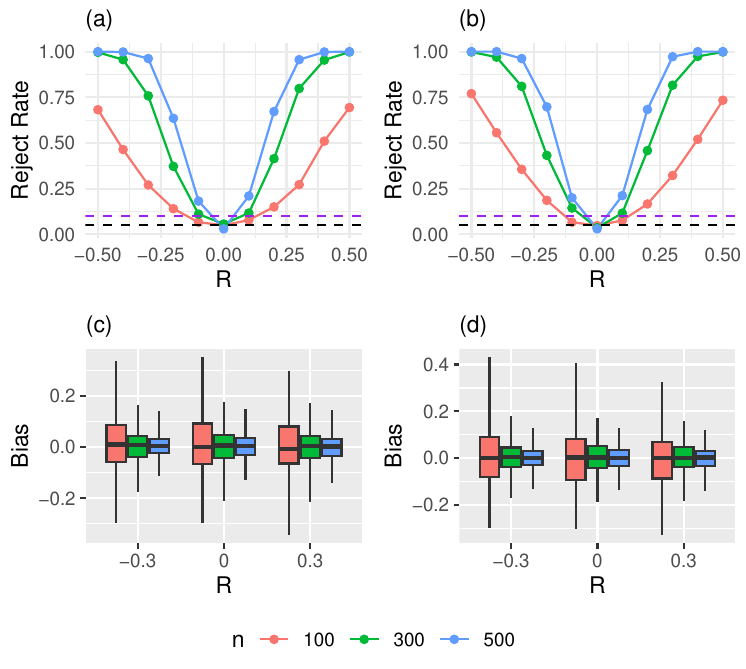}

        \end{minipage}\hfill
    \begin{minipage}{0.48\textwidth}
        {\raggedright
D \par}
        \centering
        \includegraphics[scale=0.6]{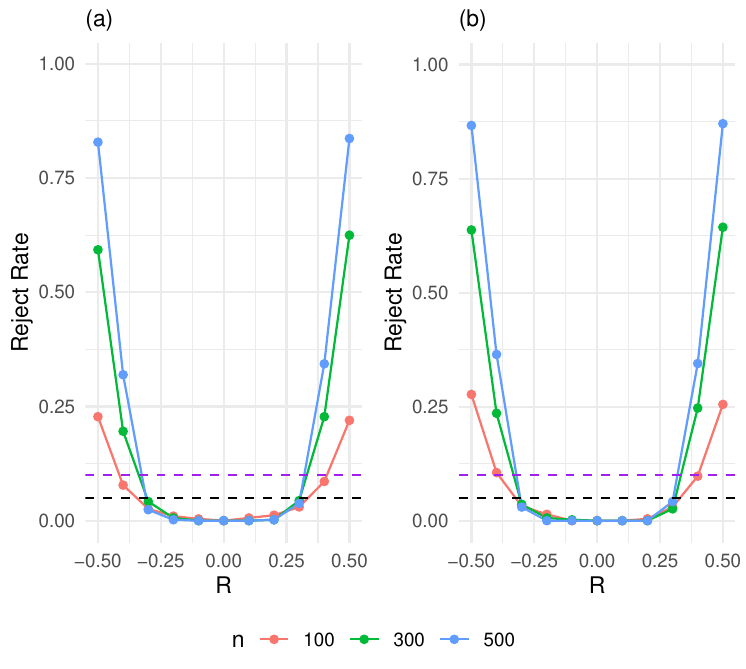}

    \end{minipage}

\caption{Sub-figure A, B, and C correspond to the linear, exponential and triangular settings, respectively.  
      Within each sub-figure, (a) and (c) correspond to $\Phi = I_p$; (b) and (d) correspond to $\Phi = DD^\intercal$; (a) and (b) show rejection rates for testing $H_0: R=0$; (c) and (d) show boxplots of the bias of the estimated microbial correlation. 
      Sub-figure D shows rejection rates for $H_0: |R| \leq 0.3$ vs. $H_0: |R| > 0.3$ under the triangular setting.
      In (a) and (b), the black line is at 0.05, and the purple line is at 0.1.}
\label{fig:combined}
\end{figure}

To evaluate the performance of our proposed method, we considered a wide range of $R_0$ from $-0.5$ to $0.5$ with a step of 0.1. 
For each simulation setting, we carried out 500 independent replications. We evaluated the estimation accuracy of $R$ based on the original estimator $\widehat{R}$ in \eqref{Rhat}
as well as the type-I error rates and power of the proposed method for testing $H_0: R = 0$ vs. $H_1: R \neq 0$ based on the calibrated estimator $\widehat{R}_{ss}$ in \eqref{Rhat:new}. 




Figure \ref{fig:combined}A-C represent the linear, exponential, and triangular settings, respectively.
From all figures, we observe that the performance of our method improves with an increase in the sample size of the PM2S study (i.e., $n$) for both estimation and inference. 
We take the linear scenario as an example.  
As shown in Fig.~\ref{fig:combined}A(a) and (b), for both scenarios of \( \Phi \), the type-I error rates are well controlled at 0.05 for all sample sizes with those in Fig.~\ref{fig:combined}A(a) controlled more stringently. 
Correspondingly,  the power in Fig.~\ref{fig:combined}A (a) is slightly lower than that in  Fig.~\ref{fig:combined}A (b). 
In terms of estimation accuracy, from Fig.~\ref{fig:combined}A(c) and (d), we observe that the median biases for all values of $R$ are around 0.
Furthermore, these biases decrease substantially as the sample size increases. 
These characteristics observed in Fig.~\ref{fig:combined}A were also observed in Figs.~\ref{fig:combined}B and C. 


In real applications, practitioners may be particularly interested in detecting strong microbial correlations between metabolites,  
which can serve as hypotheses for future research.  
 Thus, we conducted an additional analysis by testing  
\( H_0: |R| \leq 0.3 \) against \( H_1: |R| > 0.3 \)  
under the triangular setting.  
The type-I error rates and statistical power are presented in Fig.~\ref{fig:combined}D for the two scenarios of \( \Phi \).  
When the true microbial correlation \( R \in (-0.3, 0.3) \), the rejection rate is close to 0, which aligns with our definition of the test statistic in \eqref{test:R0}.  
Specifically, when \( R \in (-0.3, 0.3) \), the estimated correlation is also likely to fall within this range.  
From \eqref{test:R0} and \eqref{pval:R0}, it follows that  
if the estimated microbial correlation satisfies \( -0.3 \leq \widehat{R} \leq 0.3 \), then the p-value in \eqref{pval:R0} is exactly 1.  
While this suggests that our test statistic is conservative—potentially leading to low statistical power—Figure \ref{fig:combined}D shows that when \( |R| = 0.5 \), which is only 0.2 units away from the null hypothesis threshold \( R_{0} = 0.3 \), we still achieve relatively good power, except in the case where \( n = 100 \).  



In all simulations in this section, the external sample size \( N \) satisfies \( N = 10n \), while Condition \ref{cond:aPhi:bPhi} requires \( n = o(N^{1/4}) \).  
Thus, our simulation results suggest that, although theoretically, a substantially large external dataset is required, in practice, a smaller external sample than theoretically prescribed may still yield accurate estimations, good type-I error rate control, and reasonable statistical power.  
In the real dataset analyzed below, the external sample size is approximately 70 times the PM2S sample size, which, presumably, should provide more accurate and robust results compared to those observed in the simulations.


\section{Data Application}
\label{sec:application}
We revisited the paired metagenomic-metabolomic study (PM2S) from the Yachida cohort \citep{yachida2019metagenomic}. 
In this study, whole-genome shotgun metagenomics and capillary electrophoresis time-of-flight mass spectrometry (CE-TOFMS)-based metabolomics were employed on fecal samples collected from healthy individuals as well as patients at various stages of colorectal neoplasia. 
We obtained metagenomic and metabolomic data for 127 healthy individuals. 
Additionally, we selected 9,130 healthy subjects with metagenomic data by combining 93 studies, where the detailed information of these studies can be found in \cite{curated}. 
Potential batch effects were removed across studies using ConQuR \citep{ling2022batch}. 
Our goal is to study healthy microbial co-metabolism, using the Yachida PM2S as the target study and leveraging the 9,130 additional subjects as an external cohort. This analysis is expected to serve as a benchmark for future studies on diseased populations.




The metagenomic data is characterized by an excessive number of zeros. 
Specifically, more than $90\%$ of the data entries were zero, and $93\%$ of the species were absent in at least half of the subjects.
To address this issue, we aggregated the species within each microbial 
family
and further filtered out microbial families absent in more than 80\% of the subjects.
Consequently, we obtained family-level metagenomic data for 255 families.  
The same data processing was applied to the external cohort. 
Due to the heterogeneity across the 93 studies, only 17 microbial families were retained and 12 of them were overlapping with the Yachida cohort. 
We thus took the overlapping 12 microbial families for downstream evaluations of microbial correlations. 
To account for the compositional nature of microbiome data, 
we applied the centered log-ratio transformation separately for the Yachida cohort and the external cohort. 
The metabolomics data
in the Yachida cohort contains 450 different compounds, where 403 had
unique 403 HMDB IDs \citep{hmdb}. 
We further removed metabolites that were missing in at least 50\% of the individuals, and 160 metabolites remained for downstream analysis. 
To account for the skewness in the data, we performed a log-transformation on the metabolite levels per sample. 


We analyzed the microbial correlation between each pair of the 160 metabolites with respect to the selected 12 microbial families by adjusting for age and sex. 
 As in simulation studies, we implemented sample splitting to perform statistical inference.
 Specifically, we split the Yachida cohort into two parts separately for estimating $\boldsymbol{\beta}$ and $\boldsymbol{\gamma}$ and used the external cohort to estimate $\Phi$.
 However, we found that the resulting estimator of $R$ exhibited a varied range of values across different sample splittings, suggesting potential heterogeneity across the Yachida individuals.
 This issue has been observed and discussed for sample-splitting-based methods in the existing literature \citep{meinshausen2009p}. 
However, the same issue was not observed in simulations because simulated samples were drawn from the same distribution. 
We addressed this issue by performing 100 sample splitting and 
obtained an estimator of $\widehat{R}_{ss}$ based on each split. 
We calculated the median of these 100 estimators and found the corresponding p-value based on the median. 
Finally, recognizing the complex dependencies across metabolites, 
we applied the Benjamini–Yekutieli procedure \citep{benjamini2005false} across all 12,720 comparisons to control the false discovery rate at 0.05.  

\begin{figure}
    \centering
    \includegraphics[trim={0cm 0cm 0cm 0.5cm},clip,width=1\linewidth]{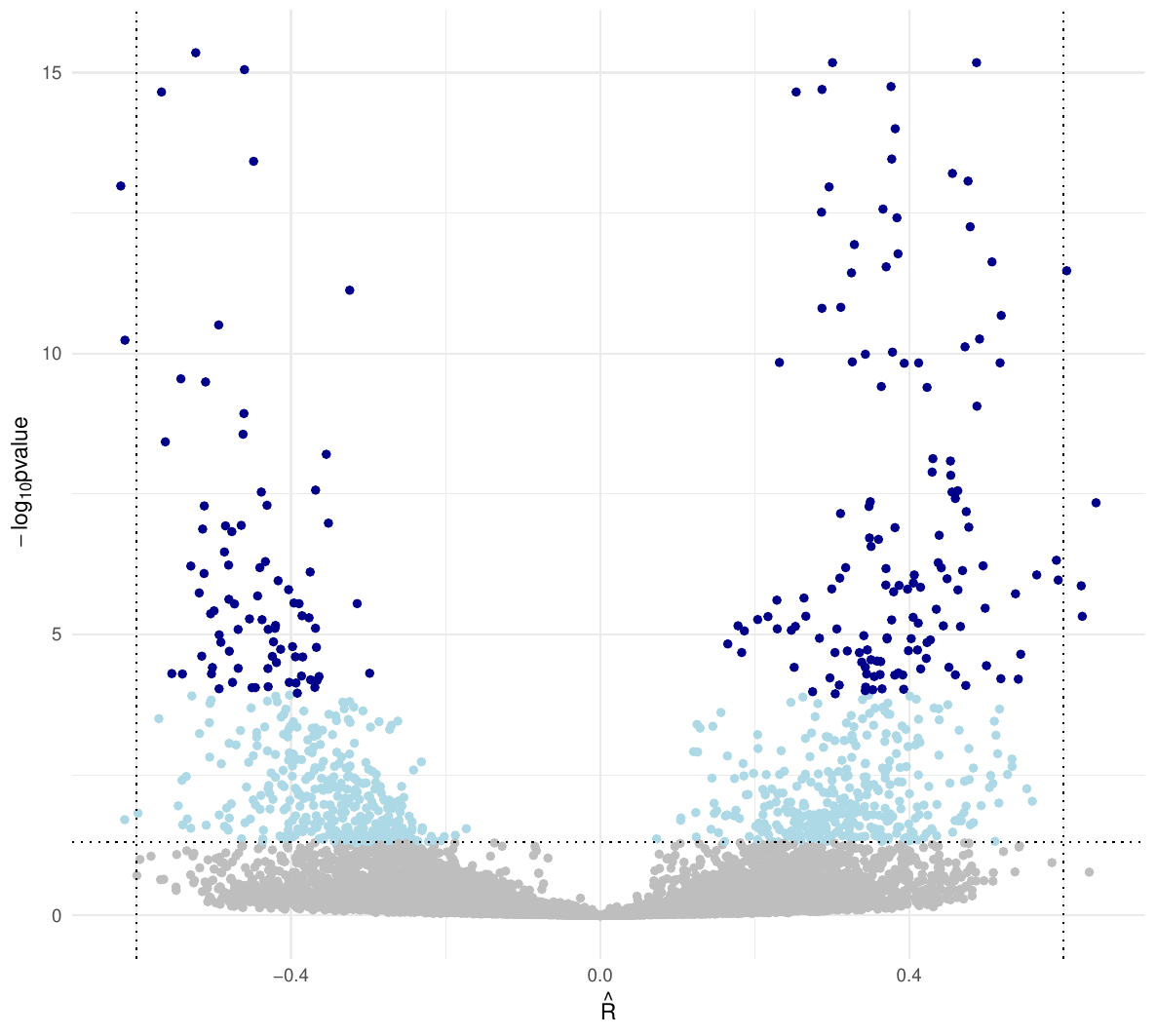}
    \caption{A volcano plot displays the median microbial correlation estimates across 100 sample splittings along with their corresponding p-values.  
Each point represents a pair of metabolites. At a significance level of 0.05, grey points indicate non-significant pairs, while light blue points denote significance before FDR control. Dark blue points indicate significance after FDR control.  
The dash-dotted line represents \( -\log_{10}(0.05) \).  
}
    \label{fig:vol}
\end{figure}

Figure \ref{fig:vol} shows a volcano plot for all pairwise microbial correlations between the 160 selected metabolites (12,720 comparisons). 
Despite the high multiple-testing burden, we were still able to detect 302 metabolite pairs displaying significant microbial correlations after FDR adjustment at 0.05. 
Among them, 205 (68$\%$) showed positive microbial correlations. 
This is consistent with the metabolome-wide Pearson correlations in Fig.  \ref{fig:heatmap}, where most of the correlations were positive. 
However, we emphasize that the microbial correlations in Fig. \ref{fig:vol} should be interpreted differentially than those Pearson correlations in Fig. \ref{fig:heatmap}. 
Our microbial correlation highlights the relationship between metabolites driven by microbes, whereas the Pearson correlation accounts for all factors, including all microbes (as shown in Fig. \ref{fig:heatmap}) and other influences such as environment, age, etc.

\begin{figure}
    \centering 
    \includegraphics[trim={3cm 2.6cm 3cm 3cm},clip,width=0.6\linewidth]{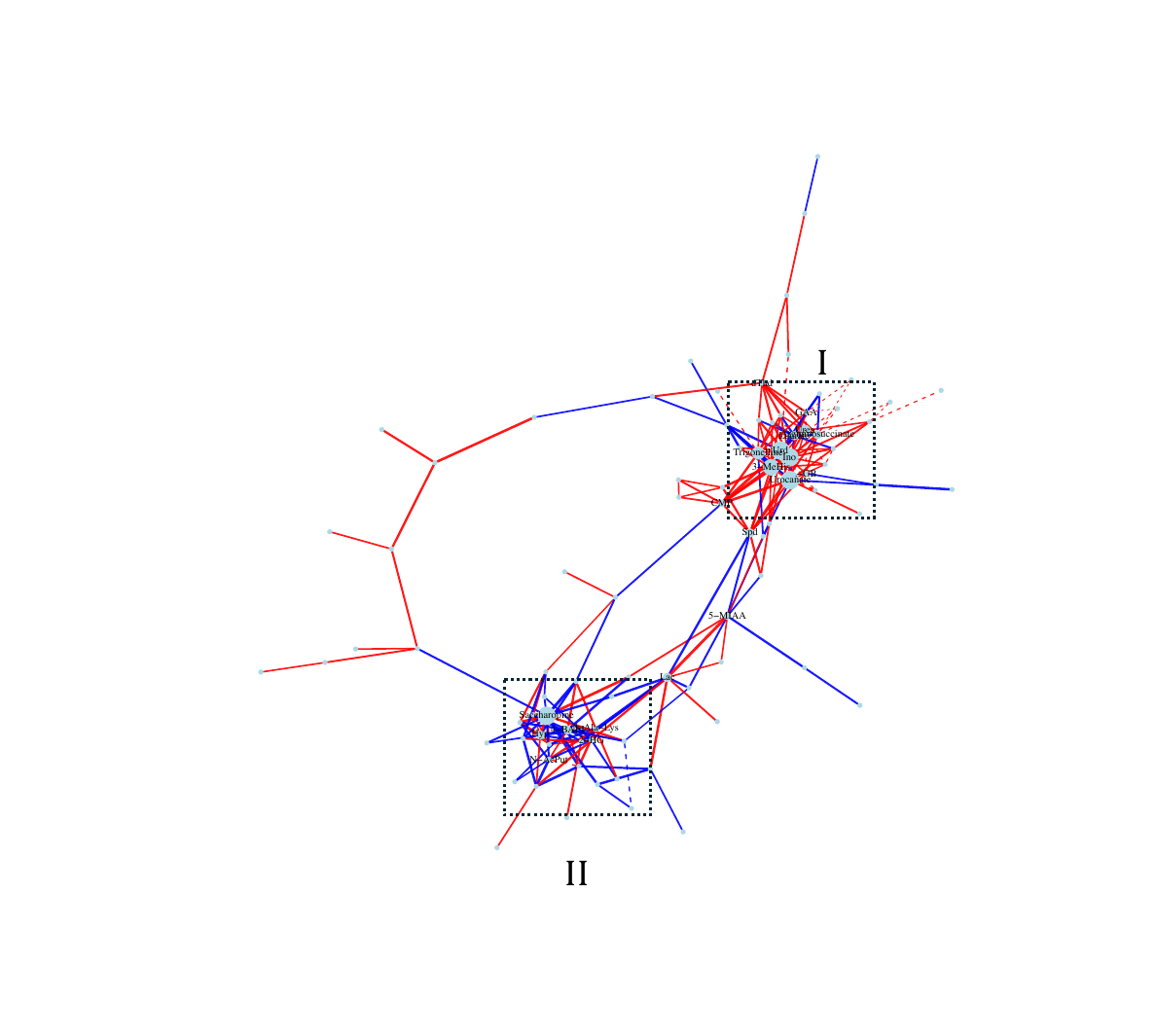}
    \caption{A microbial co-metabolism network with 160 metabolites as nodes. Edges correspond to significant microbial correlation after FDR control. 
    Red edges indicate positive correlations, while blue edges indicate negative correlations. Solid lines for $|\widehat{R}| \ge 0.3$, dashed lines for $|\widehat{R}| < 0.3$.}
    \label{fig:network}
\end{figure}
\begin{figure}

    \centering
    \begin{minipage}{0.33\textwidth}
    {\raggedright
A\par}
        \centering
        \includegraphics[width=\linewidth, 
                         trim={1.75cm 1.75cm 1.75cm 1.75cm}, 
                         clip]{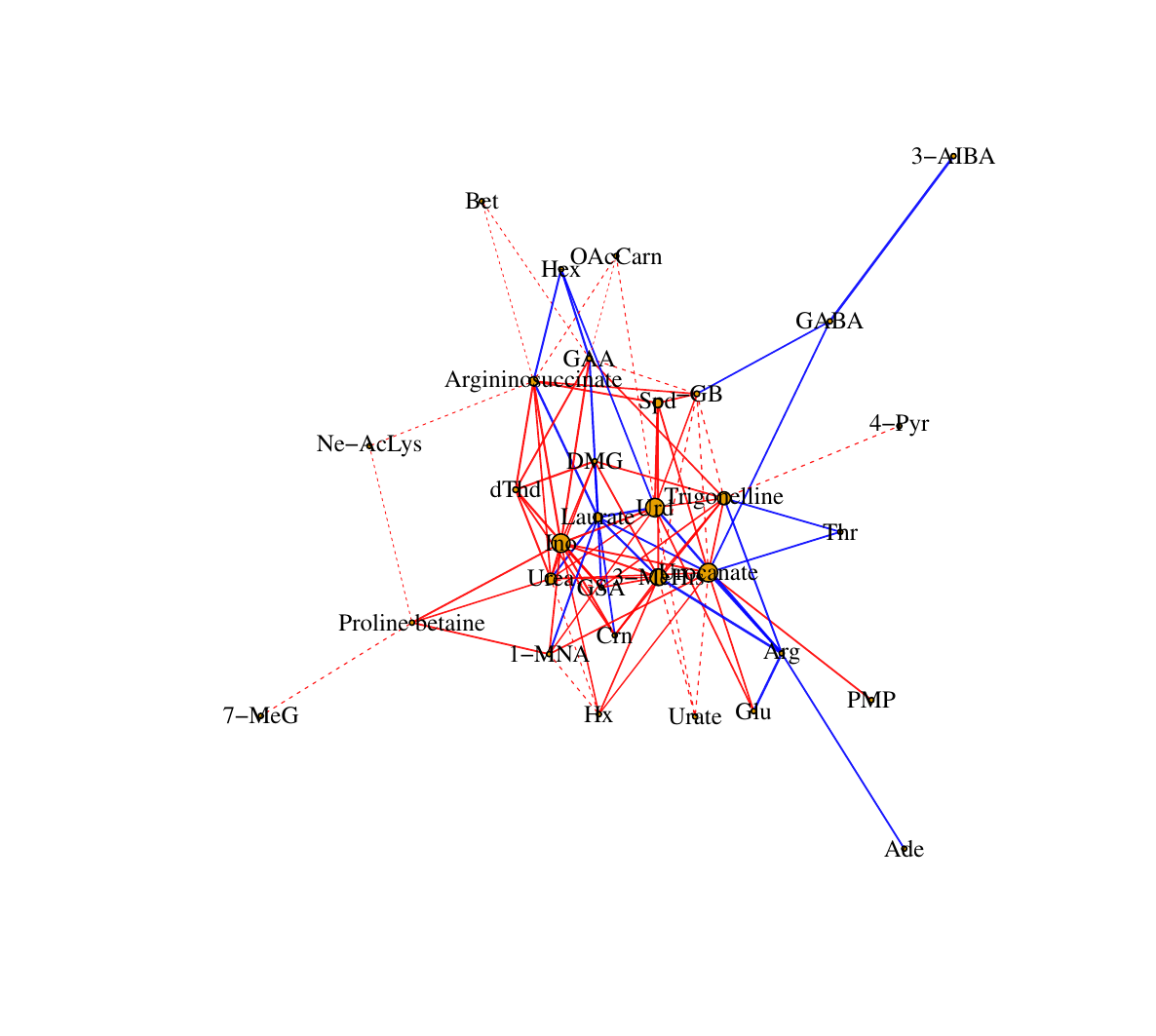}
    \end{minipage}%
    \begin{minipage}{0.33\textwidth}
{\raggedright
B\par}
        \centering
        \includegraphics[width=\linewidth,
                         trim={1.75cm 1.75cm 1.75cm 1.75cm},
                         clip]{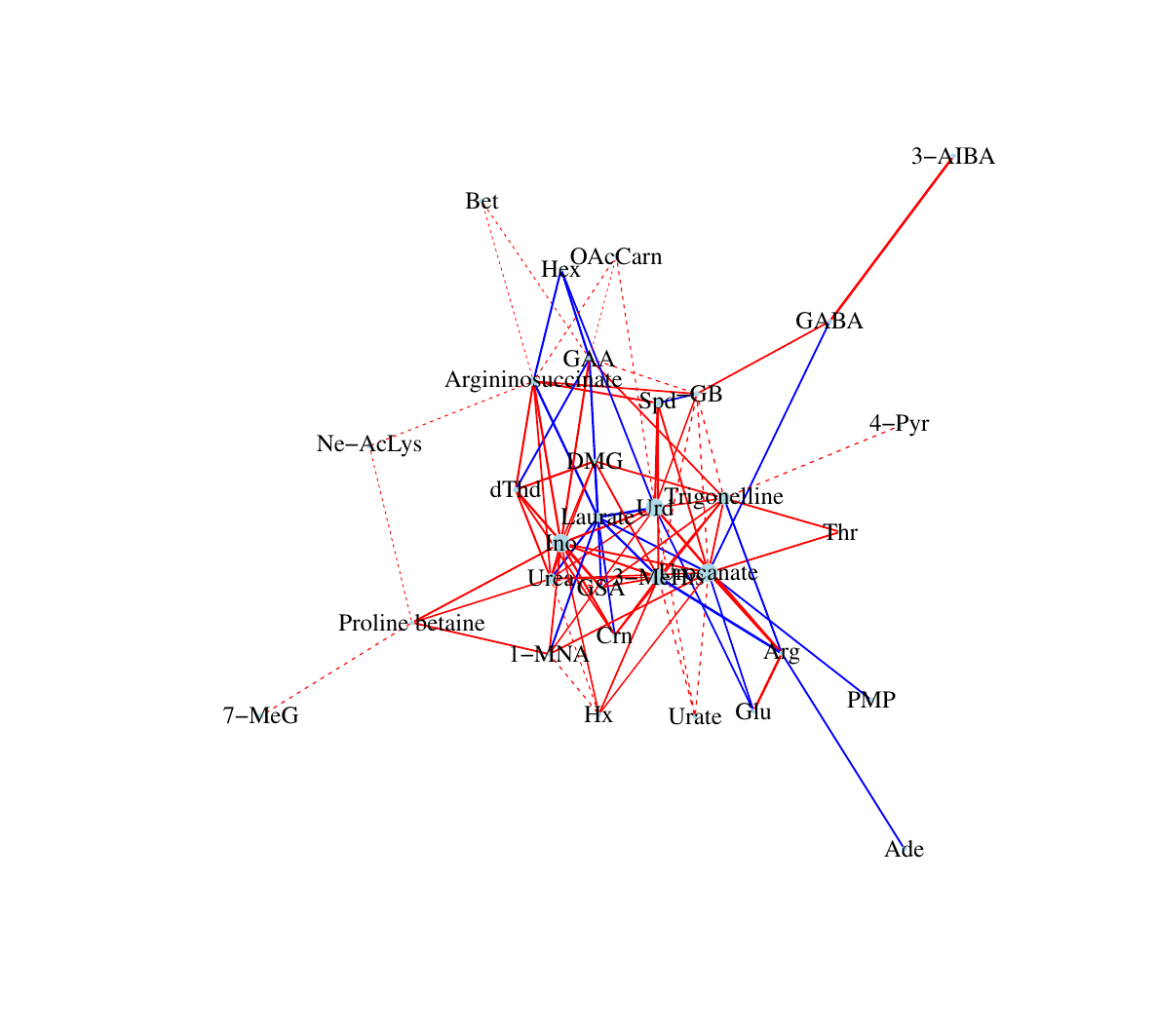}
    \end{minipage}%
    \begin{minipage}{0.33\textwidth}
    {\raggedright
C\par}
        \centering
        \includegraphics[width=\linewidth,
                         trim={1.75cm 1.75cm 1.75cm 1.75cm},
                         clip]{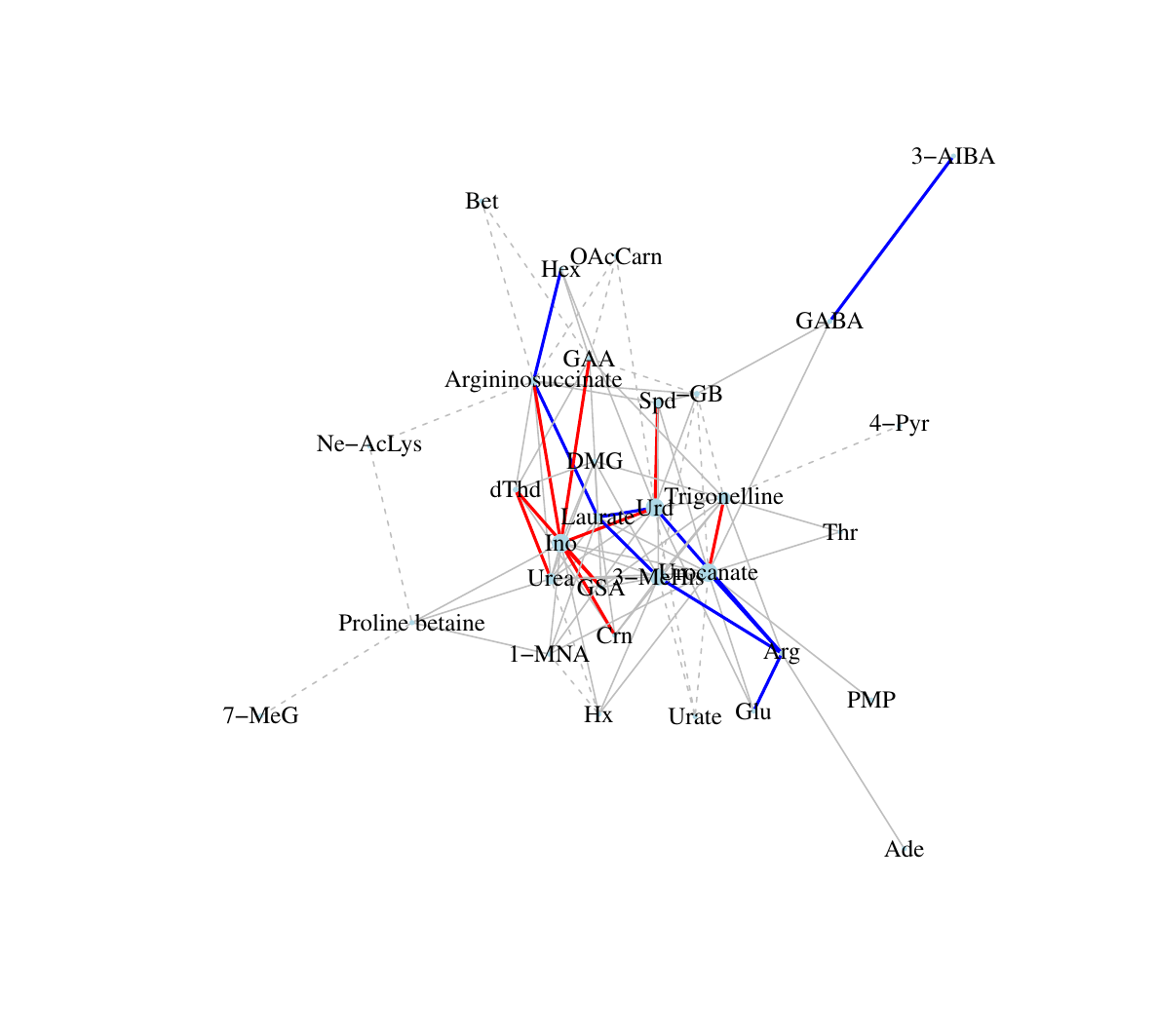}
    \end{minipage}
    \caption{ Networks of Cluster I in Fig.~\ref{fig:network}. Subfigure A presents a magnified view of Cluster I, while Subfigure B illustrates the Pearson correlations between metabolites in Cluster I.  
In both A and B, red edges indicate positive correlations, and blue edges indicate negative correlations. Solid lines represent \( |\widehat{R}| \geq 0.3 \), whereas dashed lines correspond to \( |\widehat{R}| < 0.3 \).  
Edges in Subfigure C denote significant results from testing \( H_0: |R| \leq 0.3 \) against \( H_1: |R| > 0.3 \).  
For C, gray edges indicate non-significant results.  
}
\label{fig:cluster1}
\end{figure}

We next visualized the results in Fig. \ref{fig:vol} as a microbial co-metabolism network (Fig. \ref{fig:network}), where each node represents a metabolite, edges indicate significant microbial correlations after FDR correction, and red and blue edges represent positive and negative correlations, respectively.  
Within this network, we have identified two major clusters I and II.  
Each cluster comprises metabolites that are highly interconnected based on their microbial correlation patterns.  
Cluster I mostly comprised of microbially positively correlated metabolites. 
Figure \ref{fig:cluster1}A is a magnified section of Cluster I, showing major hub nodes like urocanate, dodecanoate (laurate) and argininosuccinate. 
According to the existing literature,
they are involved in several metabolic pathways, such as the urea cycle, amino acid metabolism, or nucleic acid metabolism \citep{nelson2008lehninger}. 
 Their microbial correlations are likely driven by certain bacterial families (e.g., Enterobacteriaceae), which play critical roles in gut metabolism related to pathways such as amino acid metabolism or the urea cycle \citep{martinson2019rethinking, moreira2024enterobacteriaceae}.

\begin{figure}

    \centering
    \begin{minipage}{0.33\textwidth}
    {\raggedright
A\par}
        \centering
        \includegraphics[width=\linewidth, 
                         trim={1.15cm 1.15cm 1.15cm 1.15cm}, 
                         clip]{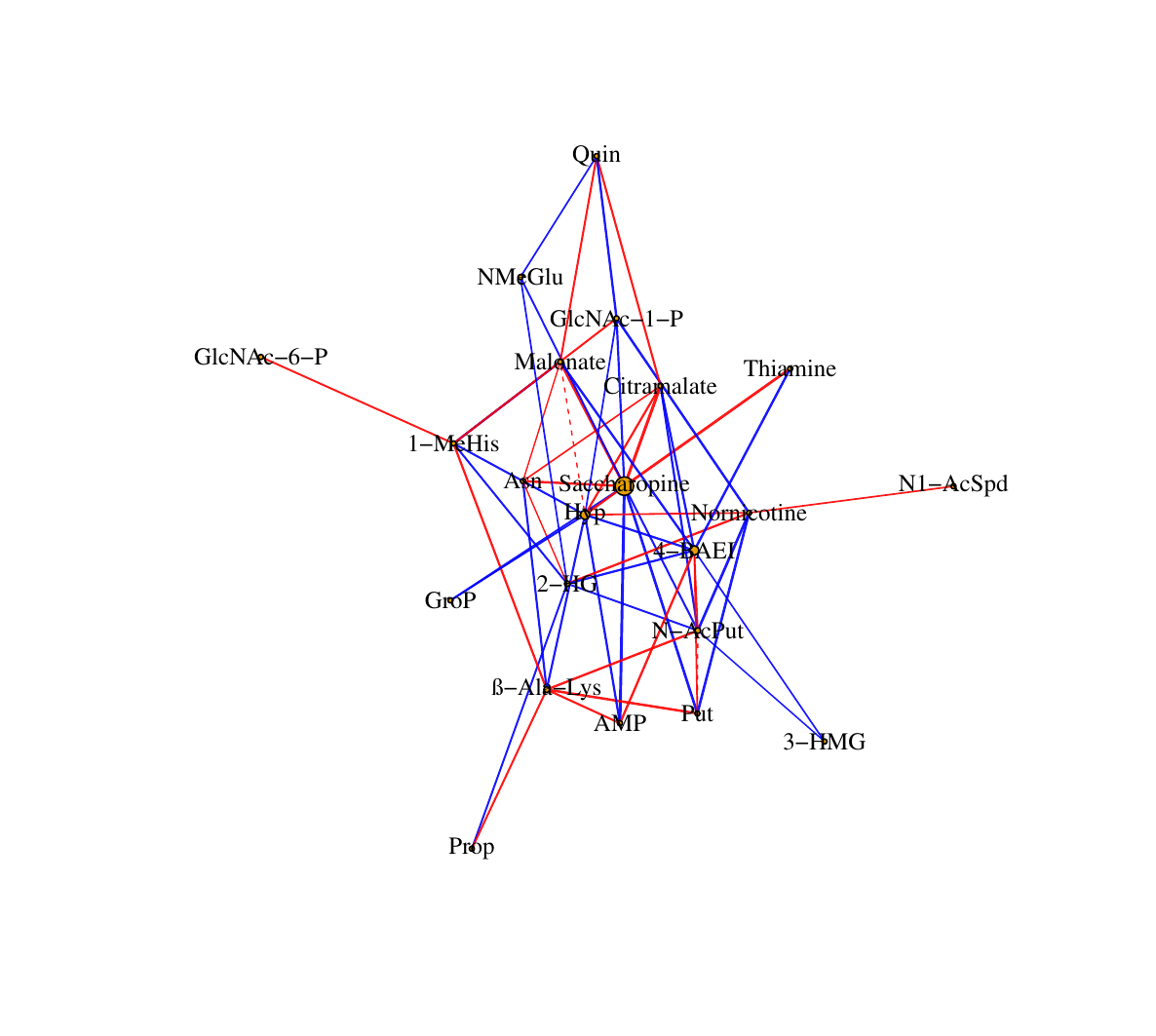}

    \end{minipage}%
    \begin{minipage}{0.33\textwidth}
    {\raggedright
B\par}
        \centering
        \includegraphics[width=\linewidth,
                         trim={1.15cm 1.15cm 1.15cm 1.15cm},
                         clip]{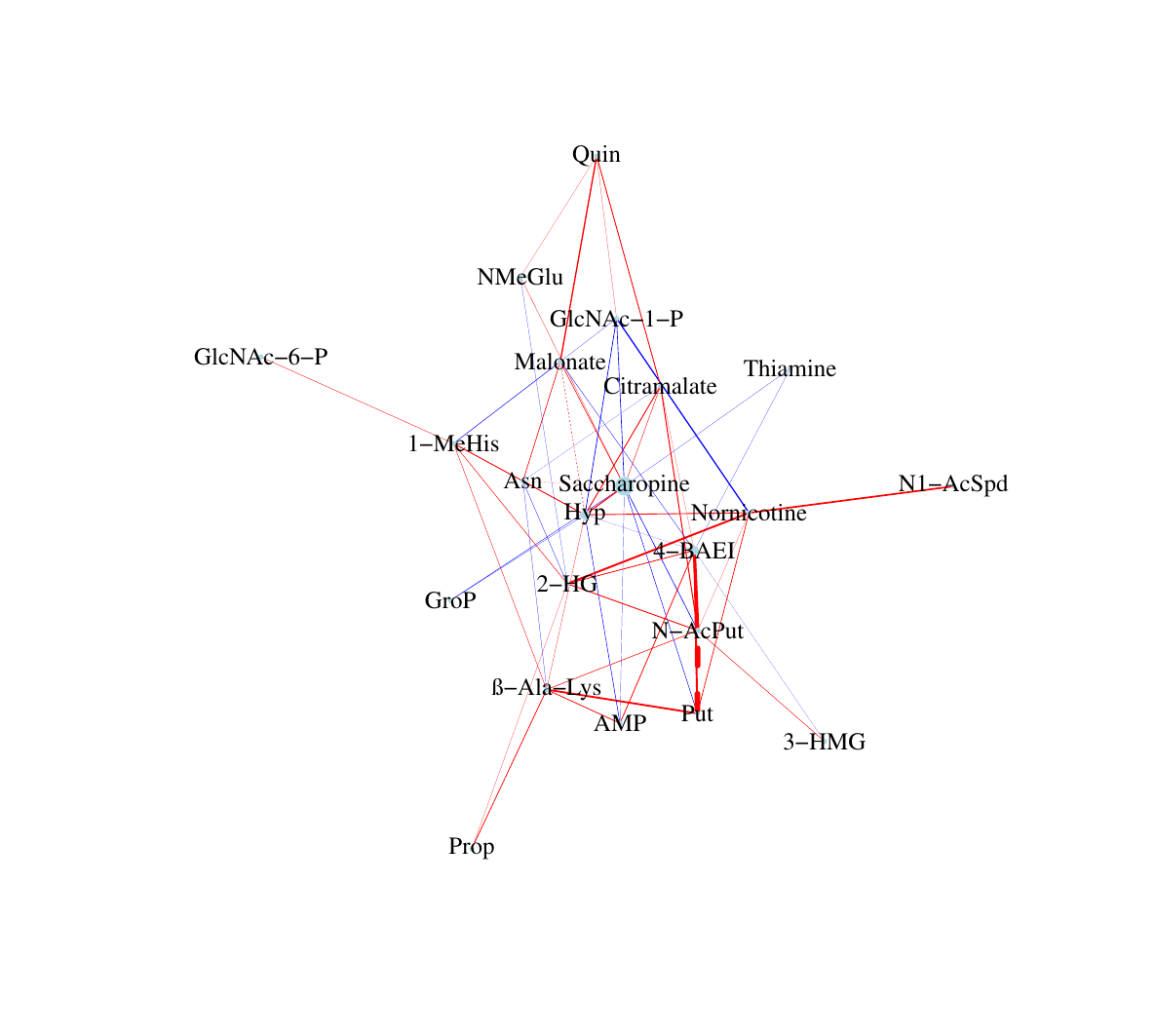}

    \end{minipage}%
    \begin{minipage}{0.33\textwidth}
    {\raggedright
C\par}
        \centering
        \includegraphics[width=\linewidth,
                         trim={1.15cm 1.15cm 1.15cm 1.15cm},
                         clip]{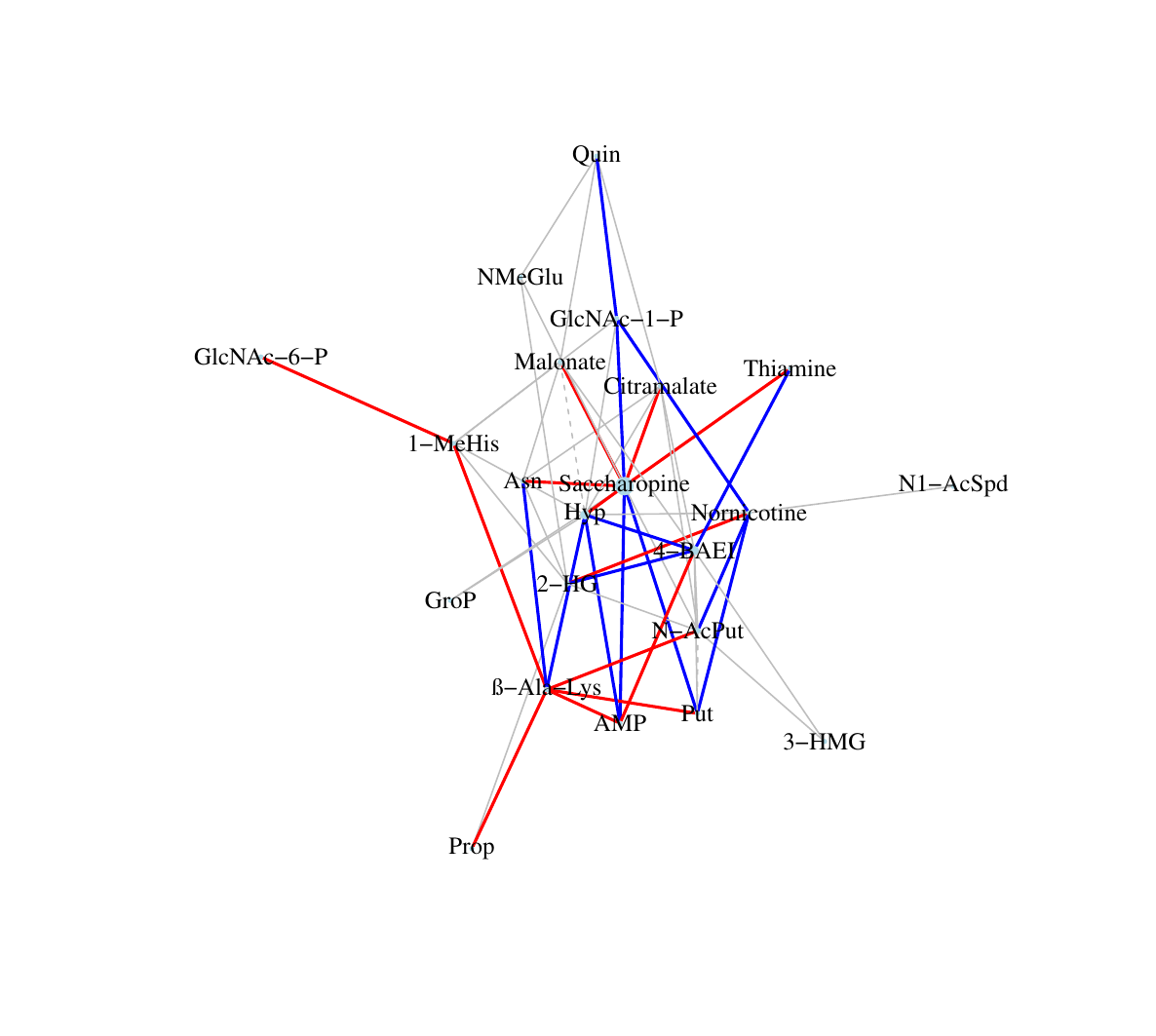}

    \end{minipage}
    \caption{Networks of Cluster II in Fig.~\ref{fig:network}. Subfigure A presents a magnified view of Cluster II, while Subfigure B illustrates the Pearson correlations between metabolites in Cluster II.  
In both A and B, red edges indicate positive correlations, and blue edges indicate negative correlations. Solid lines represent \( |\widehat{R}| \geq 0.3 \), whereas dashed lines correspond to \( |\widehat{R}| < 0.3 \).  
Edges in Subfigure C denote significant results from testing \( H_0: |R| \leq 0.3 \) against \( H_1: |R| > 0.3 \).  
For C, gray edges indicate non-significant results.  }
    
\label{fig:cluster2}
\end{figure}

Cluster II, magnified in Fig. \ref{fig:cluster2}A, mostly comprised of microbially negatively correlated metabolites. Most of these compounds, such as saccharopine, \(\beta\)-Ala–Lys, and hydroxyproline (Hyp) are intermediates or end-products of protein and amino acid catabolism \citep{10.1093/nar/gkae1059}. 
Gut microbes utilize these molecules as carbon and nitrogen sources, thereby influencing the overall structure of the microbial community \citep{wu2011proline}. 
In addition, putrescine (Put) derivatives and imidazole/histamine (4-BAEI) derivatives play key roles in modulating gut epithelial integrity, mucosal immunity, and inflammation \citep{nakamura2021symbiotic} \citep{fujisaka2018diet}. 
 Furthermore, polyamine derivatives (e.g., N-acetylputrescine) and histamine-like molecules (e.g., 4-BAEI) also act as signaling molecules between microbes and the host \citep{rojo2015clostridium}. 
 


For these two clusters, we further compared our microbial correlations with corresponding Pearson corrections, shown in Fig. \ref{fig:cluster1}B and \ref{fig:cluster2}B.
Comparing Fig. \ref{fig:cluster1}A and B suggests that 
7 pairs of negatively correlated metabolites in Fig. \ref{fig:cluster1}A have positive Pearson correlations in Fig. \ref{fig:cluster1}B, and 4 pairs show the reverse pattern. 
For example, urocanate and glutamate (Glu) exhibit a negative Pearson correlation, whereas they are positively microbially correlated.  
This positive microbial correlation is scientifically reasonable because the gut microbiota facilitates converting histidine to 
urocanate and finally to glutamate \citep{brosnan2020histidine}. 
Therefore, more urocanate would lead to more glutamate through the gut microbiome activities, 
which, however, cannot be captured by the Pearson correlation. 
Similarly, from Fig. \ref{fig:cluster2}A and B, 
14 pairs of negatively correlated metabolites in Fig.~\ref{fig:cluster2}A have positive Pearson correlations in Fig.~ \ref{fig:cluster2}B, and 4 pairs show the reverse pattern.  
For example, Propionate (Prop) and 2-Hydroxyglutarate (2-HG) show negative microbially correlation in Fig.~\ref{fig:cluster2}A but positive Pearson correlation Fig~\ref{fig:cluster2}B. 
Propionate is a well-known short-chain fatty acid (SCFA) in gut microbiota research, and 2-hydroxyglutarate is also a metabolic product of the gut microbiota \citep{ternes2022gut}. 
While no existing literature has studied their correlation,  
their negative microbial correlation may be explained by their different favored gut environment: 
The production of 2-hydroxyglutarate favors a low-oxygen environment, while Propionate requires fermentation of dietary fiber \citep{rebersek2021gut}.

We then applied the proposed microbial correlation test with \(H_0: |R|\leq 0.3\) vs. $H_1: |R| > 0.3$ to identify metabolite pairs with significant microbial correlations over 0.3. 
Figure \ref{fig:cluster1}C showed the results for Cluster I, where 15 edges remained significant. 
Inosine (Ino) serves as a central hub metabolite connected significantly to six others: guanidinoacetate (GAA), creatinine (Crn), argininosuccinate, yhymidine (dThd), guanidinosuccinate (GSA), and uridine (Urd). 
Specifically, the microbial correlation between inosine and GAA may be explained by metabolic pathways in which enhanced conversion of GAA reduces purine degradation, thereby lowering inosine levels  
\citep{mcconell2005creatine}.  
 Inosine also promotes ammonia detoxification, thereby boosting argininosuccinate flux \citep{van1992purine,guinzberg1987effect}. 
 In nucleotide metabolism, inosine and thymidine are interconnected via salvage pathways, maintaining nucleotide equilibrium \citep{wang2020inosine}. 
 Inosine and uridine, as alternative substrates in nucleotide and energy metabolism, typically vary in parallel to preserve nucleotide balance \citep{strefeler2024nucleosides}.



Similarly, Fig. \ref{fig:cluster2}C presented 24 significant edges by testing \(H_0: |R|\leq 0.3\) vs. $H_1: |R| > 0.3$ for Cluster II, where 
saccharopine plays central hub roles. 
Saccharopine is an intermediate of lysine metabolism via the saccharopine pathway. In the gut, saccharopine can be produced by microbial lysine degradation and thus influence host metabolism \cite{leandro2019saccharopine}. 
While there is no specific regulatory interaction between saccharopine and other nodes is documented, our results indicate the unrevealing microbial correlations between saccharopine and other metabolites may be a promising future research direction. 
\section{Discussions}

Recent advancements in paired microbiome-metabolome studies (PM2S) have facilitated the investigation of gut metabolites, a key mechanism through which the gut microbiome influences host health. We propose microbial correlation as a method to examine microbial co-metabolism between gut metabolites. By leveraging existing semi-parametric methodology and theory, our approach adjusts for various confounding factors, providing a more biologically meaningful assessment of microbial co-metabolism. Additionally, we integrate the PM2S with external microbiome data without paired metabolome data, enabling more efficient hypothesis testing to identify strong co-metabolism signals while maximizing the use of existing data. 
Our findings offer valuable insights into microbial co-metabolism, aligning with existing literature and generating new hypotheses for future research.

Our application maximizes the use of existing data by integrating a relatively small PM2S study with a vast amount of microbiome data to investigate healthy microbial co-metabolism.  
One limitation of our approach is that, due to heterogeneity across external datasets, only 12 microbial families are consistently observed across all datasets.  
This reflects a common challenge in microbiome studies, where differences in sequencing platforms and data preprocessing pipelines can lead to variations in detected microbial communities.  
With advancements in microbial sequencing technology, we anticipate the availability of higher-quality datasets, which can enhance microbial correlation evaluations.  
However, as the number of analyzed microbes increases, a computational challenge arises: the dimensionality of \( \bm{\beta} \) and \( \bm{\gamma} \) may become too large relative to the target sample size.  
In such cases, penalized partially linear models may be required, presenting a promising direction for future research.  
An alternative approach is to apply the current framework separately to different microbial groups to achieve higher-resolution biological insights.  
For example, we could apply the proposed framework to all genera within the {Streptococcaceae} family to gain a deeper understanding of its metabolic processes.

Our microbial correlation enables meaningful downstream analyses to further explore microbial co-metabolism. For example, by using microbial correlation as a similarity measure between gut metabolites, we can construct a weighted co-metabolism network, where nodes represent gut metabolites, edges indicate significant microbial correlations, and edge weights correspond to microbial correlation values. This approach is illustrated in Figs. \ref{fig:network}, \ref{fig:cluster1}, and \ref{fig:cluster2}.
The graph Laplacian of this network can be used for spectral clustering to identify groups of gut metabolites that are co-regulated by the same microbial pathway.
This approach can be further extended using graph-informed principal component analysis (PCA) to cluster individuals with similar microbial co-metabolism patterns.
Specifically, let $\mathcal{L} \in \mathbb{R}^{p \times p}$ denote the graph Laplacian derived from the co-metabolism network and $Z \in \mathbb{R}^{n \times p}$ the sample-by-metabolite data matrix. 
The kernel matrix for the samples can then be defined as $K = Z \mathcal{L} Z^\intercal$, with clustering performed using the top eigenvectors of $K$. 

The identified microbial co-metabolism patterns can also be linked to clinical outcomes to explore their associations with host health.
Specifically, let $\mathcal{G}_1, \ldots, \mathcal{G}_M$ denote the $M$ clusters of gut metabolites identified from the co-metabolism network, and partition the gut metabolite matrix $Z$ into $Z_1, \ldots, Z_M$ accordingly. 
Performing PCA for each $Z_m$ for $m=1, \ldots, M$, we extract the top eigenvectors for each cluster, representing distinct microbial co-metabolism modules. These modules are then analyzed in a regression framework to assess their associations with the clinical outcome, identifying significant modules. Since each module corresponds to a specific microbial co-metabolism pattern, this approach allows us to prioritize key microbial co-metabolism patterns relevant to host health, enhancing our understanding of gut microbiome-host interactions through metabolism.





 \bibliographystyle{chicago} 
 \bibliography{ref.bib, reference_R21_1, reference_R21_2, reference_R21_3}

\end{document}